\documentclass[a4paper,11pt]{article}
\usepackage{jcappub} 

\usepackage{mathptmx}       
\usepackage{helvet}         
\usepackage{courier}        
\usepackage{type1cm}        

\usepackage{makeidx}         
\usepackage{graphicx}        
\usepackage{mathtools}
\usepackage{multirow}
\usepackage{multicol}        
\usepackage{float}
\usepackage{array}
\usepackage[bottom]{footmisc}

\newcommand{\be}{\begin{equation}}
\newcommand{\ee}{\end{equation}}
\newcommand{\bea}{\begin{eqnarray}}
\newcommand{\eea}{\end{eqnarray}}

\newcommand{\der}{\ensuremath{{\rm d}}}

\title{Constraining the dark energy statefinder hierarchy in a kinematic approach}
\author{Ankan Mukherjee \footnote{E-mail: ankanju@iisermohali.ac.in},
Niladri Paul \footnote{E-mail: npaul@iucaa.in},
H. K. Jassal \footnote{E-mail: hkjassal@iisermohali.ac.in}}
\affiliation{{\em $^{1,3}$ Department of Physical Sciences,~~\\Indian Institute of Science Education and Research Mohali,\\ Sector 81, Mohali, Punjab 140306, India.}\\[2mm]
{\em $^{2}$ Inter-University Centre for Astronomy and Astrophysics,\\ Ganeshkhind, Post Bag 4, Pune 411007, India.}\\[15mm]}
\abstract{In the present work, we have adopted a kinematic approach for constraining the extended null diagnostic of concordance cosmology, known as the {\it statefinder  hierarchy}. A Taylor series expansion of the Hubble parameter has been utilised for the reconstruction. The coefficients of the Taylor series expansion are related to the kinematical parameters like the deceleration parameter, cosmological jerk parameter etc. The present values of the kinematical parameters are constrained from the estimated values of those series coefficients. A Markov chain Monte Carlo analysis has been carried out using the observational measurements of Hubble parameter at different redshifts, the distance modulus data of type Ia supernovae and baryon acoustic oscillation data to estimate the coefficient of series expansion of the Hubble parameter. The parameters in the statefinder diagnostic are related to the kinematical parameters. The statefinder diagnostic can form sets of hierarchy according to the order of the kinematical parameters.  The present values of statefinder parameters have been constrained. The first set in the statefinder hierarchy allows $\Lambda$CDM to be well within the 1-$\sigma$ confidence region, whereas the second set is in disagreement with the corresponding $\Lambda$CDM values at more than 1-$\sigma$ level. Another dark energy diagnostic, namely the $Om$-parameters, is found to be consistent with concordance cosmology.} 
\begin{document}
\maketitle
\flushbottom

\section{Introduction}
Observations suggest that at present the universe is undergoing a phase of accelerated expansion which started in recent past \cite{snia1,snia2, Riess:2004nr}. There are two distinct theoretical prescriptions in the literature to explain the cosmic acceleration. The way to explain cosmic acceleration within the region of General Relativity (GR) is to assume the existence of an exotic component in the energy budget of the universe. The other way to look for a possible explanation of this accelerated expansion is to find out a theory of gravity beyond GR. The exotic component introduced in the energy budget of the universe to explain the phenomenon of cosmic acceleration is called {\it dark energy}. Due to a lack of fundamental understanding about a feasible candidate of dark energy, various phenomenological prescriptions are proposed and tested with the help of different cosmological observations.

The simplest description of dark energy is the {\it cosmological constant} where the constant vacuum energy density is assumed to drive the accelerated expansion. The $\Lambda$CDM (cosmological constant ($\Lambda$) along with cold dark matter (CDM)) is the simplest prescribed model of late-time cosmology. Although it suffers from the problem of fine-tuning (the discrepancy between the observationally estimated value of $\Lambda$ and the value calculated from the quantum field theory) and the cosmic coincidence problem, the model is physically well motivated and it is in good agreement with most of the cosmological observations till date. Different aspects of the cosmological constant model of dark energy have been exhaustively discussed in \cite{Carroll:2000fy, Padmanabhan:2002ji}. However, it can not be concluded whether dark energy is a constant or evolving with time. Observations are also well fitted with different time-evolving dark energy models. Another intriguing fact is that the present observation is not enough to confirm whether the dark matter is truly cold or it also has some thermal energy. Hence the search for viable alternatives of $\Lambda$CDM model is highly relevant in the context of dark energy research. Plenty of alternative prescriptions about dark energy are there in the literature, though none of them has been declared to be the perfect one.

Reconstruction of the cosmological model is a reverse engineering technique to figure out a consistent model directly from observational data based on some phenomenological assumptions. Comprehensive reviews on different aspects of reconstruction of dark energy models are there in \cite{Sahni:1999gb,Sahni:2006pa,Peebles:2002gy,Copeland:2006wr}. There are two different approaches to the reconstruction; the first one is a parametric approach, and the other one is a non-parametric approach. The parametric approach to the reconstruction is based on some assumed parametric form of a cosmological quantity. The parametrisation can be based on a specific dark energy model or it can be model independent. Some recent attempts to constrain cosmological parameters for specific dark energy scenario are referred in \cite{Chen:2016eyp,Park:2018fxx,Durrive:2018quo,Ryan:2018aif}. Non-parametric approaches to the reconstruction have also been emphasised in this context to avoid the possibility of bias in the result. Various statistical prescriptions are there in the literature to reconstruct the cosmological dynamics in non-parametric ways. For instance, {\it principle component analysis} has been adopted by Crittenden {\it et al} \cite{Crittenden:2005wj}, Clarkson and Zunckel \cite{Clarkson:2010bm}, Ishida and Souza \cite{Ishida:pca}, Amendola {\it et al} \cite{Amendola:pca}, Qin {\it et al} \cite{Qin:2015eda}, {\it Gaussian process} has been adopted for reconstructing the dark energy equation of state parameter by Holsclaw {\it al.}\cite{Holsclaw:gp}, Seikel, Clarkson and Smith \cite{Seikel:2012uu}, Nair,Jhingan and Jain \cite{Nair:2013sna}. Recently Shafieloo, Kim and Linder have adopted the Gaussian process technique to constrain the cosmographical parameters in a model-independent way \cite{Shafieloo:2012ht}. However, this non-parametric approach to the reconstruction suffers from the lack of sufficient and suitable data sets. The uncertainty associated with the reconstructed quantities increases as one goes towards reconstructing higher order derivative of the scale factor.  The reconstructed evolutions of the Hubble parameter and the deceleration parameter are highly degenerate for different dark energy models. One prime endeavour of the present work is to constrain the higher-order kinematic terms and to invade the degeneracy in different dark energy models.

A kinematic approach towards reconstruction is independent of any prior assumption about the theory of gravity or the nature of dark energy. The only assumptions are the homogeneity and isotropy of the universe at the cosmological scale. A kinematical parameter contains only the scale factor and its time derivatives. Reconstruction of kinematical quantities can either be parametric or non-parametric. A popular kinematic approach to constrain the cosmological evolution is {\it Cosmography} \cite{weinberg,harrison,Bernstein:2003es,Visser:2004bf,Dabrowski:2004hx,Dabrowski:2005fg,Cattoen:2007sk,Aviles:2012ay,Dunsby:2015ers,Busti:2015xqa}. The idea is to express the luminosity distance in terms of the kinematical quantities which are constructed from the scale factor and its time-derivatives. Dunajski and Gibbons \cite{Dunajski:2008tg} have discussed the constraints on kinematical quantities like the deceleration parameter, cosmological jerk parameter, snap parameters and so forth for different dark energy scenarios. Parametrisation of kinematical quantities are discussed by Rapetti {\it et al.}  \cite{Rapetti:2006fv}, Zhai {\it et al.}\cite{Zhai:2013fxa}, Mukherjee and Banerjee \cite{Mukherjee:2016shl,Mukherjee:2016trt}. Non-parametric reconstruction of kinematical quantity has been discussed by Shafieloo, Kim and Linder \cite{Shafieloo:2012ht}.  Recently Balcerzak and Dabrowski have discussed the statefinder luminosity distance relation in varying speed of light cosmology \cite{Balcerzak:2014rga}.

The present work is based on a purely kinematic approach to estimate the kinematical parameters as well as the set of null diagnostics of dark energy. The starting point is a Taylor series expansion of the Hubble parameter prescribed by Aviles, Klapp and Luongo \cite{Aviles:2016wel}. In \cite{Aviles:2016wel}, the authors have looked for an unbiased way to estimate the cosmographical parameters. They have termed this method as {\it Eis} because the coefficients of the Taylor expansion of the Hubble parameter are written as $E_i$-s. The analysis has been carried out with the simulated supernova catalogues to compare the bias parameter in {\it Eis} method and in {\it cosmography}. The Union 2.1 compilation and the Joint Lightcurve Analysis (JLA) data were introduced to estimate cosmographical parameters. The present analysis has been generalised by introducing the observational measurements of the Hubble parameter at different redshifts and baryon acoustic oscillation (BAO) data along with the distance modulus data-set of Joint Lightcurve Analysis (JLA). The prime focus of the present work is not only to constrain the cosmographical parameters in a model-independent way but also to estimate the values of different dark energy diagnostics, namely the hierarchy of the {\it statefinder} diagnostic \cite{Alam:2003sc} and the {\it Om}-diagnostic \cite{Sahni:2008xx, Zunckel:2008ti}.

The dark energy diagnostics are prescribed to compare any dark energy model with concordance cosmology, namely the $\Lambda$CDM. The idea is to check whether a model, which is in good agreement with observations, always mimic the $\Lambda$CDM or behave in different ways at the higher order geometric terms. The evolution of the Hubble parameter and the deceleration parameter are well constrained from present observations. All the viable dark energy models are highly degenerate up to the second order time derivative term of the scale factor, namely the deceleration parameter. This degeneracy among different dark energy models can be broken by introducing the higher order kinematical terms. However, the higher order kinematical terms are not well constrained by present observations.  In the present work, the jerk parameter and the snap parameter, which are the dimensionless representation of the third and the fourth order time derivative of the scale factor respectively, have been emphasised and their present values are obtained based on the Taylor series expansion of the Hubble parameter, namely the {\it Eis} method \cite{Aviles:2016wel}. Recently a similar approach has been adopted by Capozziello {\it et al.} \cite{Capozziello:2018jya} to constrain the nature and evolution of dark energy in a model independent way. A model-independent assessment of late-time cosmology using Gaussian Process has been discussed recently by Haridasu {\it et al.} \cite{Haridasu:2018gqm}.

For a direct comparison with $\Lambda$CDM, we estimated the present value of the dark energy diagnostics, namely the {\it statefinder} and the $Om$. The {\it statefinder diagnostic}, introduced by Alam {\it et al} \cite{Alam:2003sc} is a null diagnostics of $\Lambda$CDM cosmology as the $\Lambda$CDM model corresponds to the $(1,0)$ point on the 2D parameter space of the statefinders. In \cite{Alam:2003sc}, the authors have prescribed the statefinder up to third order terms that incorporates the jerk parameter. Further generalization of the statefinder has been discussed by Arabsalmani and Sahni \cite{Arabsalmani:2011fz} where the authors have introduced a hierarchy of the set of statefinders. In the present work, we attempted to constrain the set of statefinder hierarchy up to the fourth order terms which include the snap parameter. It is important to note that the constraints on the set of the statefinder hierarchy obtained in the present analysis are from the kinematic approach, based on the series expansion of the Hubble parameter. Thus the estimation of the parameter values in the present analysis is not done in a non-parametric manner. However, it is independent of any assumption about the dark energy model.   

In the next section (section \ref{reconst}), the kinematic approach, adopted in the present analysis, has been discussed. In section \ref{datamethod}, the observational data and the methodology have been briefly stated as well as the constraints on the coefficients of the Taylor expansion of the Hubble parameter are presented. Constraints on the kinematical parameters like the deceleration parameter, jerk and snap parameter, obtained in the present analysis,  are reported in section \ref{Cosmographicalpar}. In section \ref{stateOm}, the constraints on the statefinder hierarchy and the $Om$ diagnostics are presented. Finally, it has been concluded with an overall discussion about the results in section \ref{conclusion}.
%
\section{Kinematic approach to the reconstruction}
\label{reconst}
The mathematical formulation of cosmology begins with the assumption of homogeneity and isotropy of the universe at the cosmological scale for which the metric is written as,
\be
\der s^2 = -\der t^2 + a^2(t) \left[\frac{\der r^2}{1-kr^2}+r^2 \der \theta^2+r^2sin^2\theta \der \phi^2\right] \,\, ,
\ee 
where we have assumed the speed of light in vacuum, $c$ to be unity. The coefficient of the spatial part of the metric $a(t)$, which corresponds to the time evolution of the spatial separation between two points, is called the {\it scale factor} and $k$ is the curvature parameter. This metric is the Friedmann-Lemaitre-Robertson-Walker (FLRW) metric. Incorporating the FLRW metric in Einstein's field equations yields the Friedmann equations  as
\be 
3\frac{\dot{a}^2+k}{a^2}=8\pi G\rho,
\label{fried1}
\ee 
\be 
2\frac{\ddot{a}}{a}+\frac{\dot{a}^2+k}{a^2}=-8\pi Gp.
\label{fried2}
\ee
The right-hand sides of these equations are obtained from the energy-momentum tensor corresponding to the components of the energy budget of the universe. These $\rho$ and $p$ are respectively the total energy density and the pressure of the components. In a kinematic approach, the model is reconstructed from the quantities which are functions of the scale factor and its time derivatives, for instance, the Hubble parameter, the deceleration parameter, the cosmological jerk parameter etc. The advantage of a kinematic approach lies in the fact that there is hardly any prior assumption about the nature of the components in the matter sector. This approach is also independent of any prior assumption about the theory of gravity. In the present analysis, we have expanded the Hubble parameter $(H)$, defined as $H=\frac{\dot{a}}{a}$, in a Taylor series. The Hubble parameter can also be presented as a function of the redshift $z=-1+\frac{a_0}{a}$, where $a_0$ is the present value of the scale factor. For all the numerical computations carried out in this work, we have taken $a_0$ to be unity. The Taylor expansion of the Hubble parameter with $z$ as the argument can be written as,  
\be
E(z) \equiv \frac{H(z)}{H_0}=\sum_{i=0}^{3}\frac{1}{i!}E_iz^i,
\label{hubexp}
\ee
where $E_i \equiv \frac{\der ^iE(z)}{\der z^i}|_{z=0}$ and $H_0$ is the value of the Hubble parameter at the present epoch. It is important to note here that only up to $3^{rd}$ order term has been taken into account in this expansion because the highest power of redshift $z$ allowed in the first Friedmann equation (equation (\ref{fried1})) is $(1+z)^6$, which corresponds to the stiff matter. The constraint $E(z=0)=1$ gives $E_0=1$. The estimation of the coefficients of the Taylor expansion ($E_1, E_2, E_3$) and the present value of the Hubble parameter ($H_0$) is important to estimate the present values of different cosmographical parameters. It is already mentioned that this is independent of any prior assumption about the distribution in the matter sector and the dark energy model. Hence the values of kinematical parameters and statefinders can be used to check the viability of different dark energy models. In a recent analysis by Aviles, Klapp and Luongo, \cite{Aviles:2016wel}, it has been shown that the series expansion method of Hubble parameter is unbiased in the estimation of the {\it cosmographical parameters}. In the present analysis, the set of parameters ($h,E_1, E_2, E_3$, $\Omega_k$), where $h= H_0/(100 ~ \rm km ~\rm sec^{-1} \rm Mpc^{-1})$ and the curvature contribution $\Omega_k=-k/H_0^2$, are constrained. Three different observational data sets, namely the observational measurements of Hubble parameter (OHD), the distance modulus measurements of type Ia supernovae from the Joint Light-curve Analysis (JLA) sample and the measurements of baryon acoustic oscillation (BAO), have been utilised for constraining the parameters.  
%
%
\section{Data, methodology and results}
\label{datamethod}
In the present analysis, only the low redshift observational data sets, namely the distance modulus measurement of type Ia supernovae (SNe), observational measurements of Hubble parameter (OHD) and the baryon acoustic oscillation (BAO) data have bee taken into account. These low redshift observations are independent of any fiducial assumption about the background cosmological model. Thus these observations are suitable for any model independent estimation of cosmological parameters. \\ 

Supernovae type Ia are the first observational candidates which indicated the accelerated expansion of the universe. Here, we use the measurement of the distance modulus $\mu(z)$ of supernova type Ia, which is the difference between its apparent magnitude ($m_B$) and its absolute magnitude ($M_B$) in the B-band of the observed spectrum. The distance modulus is defined as,
\begin{equation}
\mu(z)=5\log_{10}{\Bigg(\frac{d_L(z)}{1 \rm Mpc}\Bigg)}+25 \, .
\end{equation}
In the above expression, $d_L(z)$ is the luminosity distance of a type Ia supernova observed at redshift $z$. For a FLRW universe with a generic spatial curvature density  $\Omega_k$, the expression for luminosity distance is,
\begin{align}
d_L(z) = \frac{(1+z)}{H_0} Re \left[ \frac{\sinh (\sqrt{\Omega_k}\chi (z)H_0)}{\sqrt{\Omega_k}} \right] \,\, , \label{eq:lum_dis_dis_general}
\end{align} 
where $\chi(z) = \int_0^z \der z^{\prime}/H(z^{\prime})$.

In the current work, we use 31 binned distance modulus data sample of the recent Joint Light-curve Analysis (JLA) \cite{jla}. The observational measurements of the Hubble parameter (OHD) at different redshifts in the range $0.07<z<2.36$ by different groups, have also been used in the present analysis to constrain the model parameters.  This OHD data points, utilised in the present analysis, are mainly measured by three different methods, Cosmic Chronometer method \cite{ohdcc}, BAO signals in the galaxy distribution \cite{ohdbao} and the BAO signal in Lymann-$\alpha$ forest distribution \cite{ohdLya}. In the present analysis, the BAO measurements by 6dF Galaxy Survey at redshift $z=0.106$ \cite{6dF}, the Sloan Digital Sky Survey (SDSS) Data Release 7 (DR7) Main Galaxy Sample at redshift $z=0.15$ \cite{Ross:2014qpa},  the Baryon Oscillation Spectroscopic Survey (BOSS) data at redshift  $z=0.32$ (BOSS LOWZ) and $z=0.57$ (BOSS CMASS) \cite{Anderson:2013zyy} have also been utilized. We have kept $r_s$, the acoustic sound horizon at photon drag epoch, as a free parameter in our analysis. \\

The statistical analysis has been carried out with Bayesian inference technique where the {\it posterior} probability distribution function of the model parameters is proportional to the {\it likelihood} function and the {\it prior} information about the probability distribution of the parameters, i.e. $posterior\sim likelihood \times prior$. The likelihood function, ${\mathcal L}(\{\theta\})$ is defined as,
\be
- \ln {\mathcal L}(\{\theta\}) = \frac{1}{2}(\mu^*-\mu)^{T} {\bf C}_{\mu}^{-1} (\mu^* - \mu) + \sum_{i=1}^N\frac{(\eta^*(z_i)-\eta(z_i,\{\theta\}))^2}{2\sigma_i^2},
\label{likeli}
\ee
where $\{\theta\}$ is the set of parameters, $\bf{\mu}$ denotes the vector associated with distance moduli and $\bf{C_{\mu}}$ is the full covariance matrix associated with the binning of distance moduli measurements as given in \cite{jla}. The other quantities $\eta (z_i)$ are associated with either OHD or BAO measurement at redshift $z_i$ with $\sigma_i$ being the corresponding error in the measurement. The total number of data-points in the OHD + BAO data set has been denoted as $N$ and the quantities with asterisk ($*$) are the observational measurements.  In the present analysis, a uniform prior has been assumed for the parameters. The parameter estimation has been done in Markov chain Monte Carlo (MCMC) method using the {\footnotesize PYTHON} implementation of a MCMC sampler, namely the {\footnotesize EMCEE}, introduced by Goodman and Weare \cite{emcee1} and by Foreman-Mackey {\it et al} \cite{emcee}.
\begin{figure}[tb]
\begin{center}
\includegraphics[width = 0.8\textwidth]{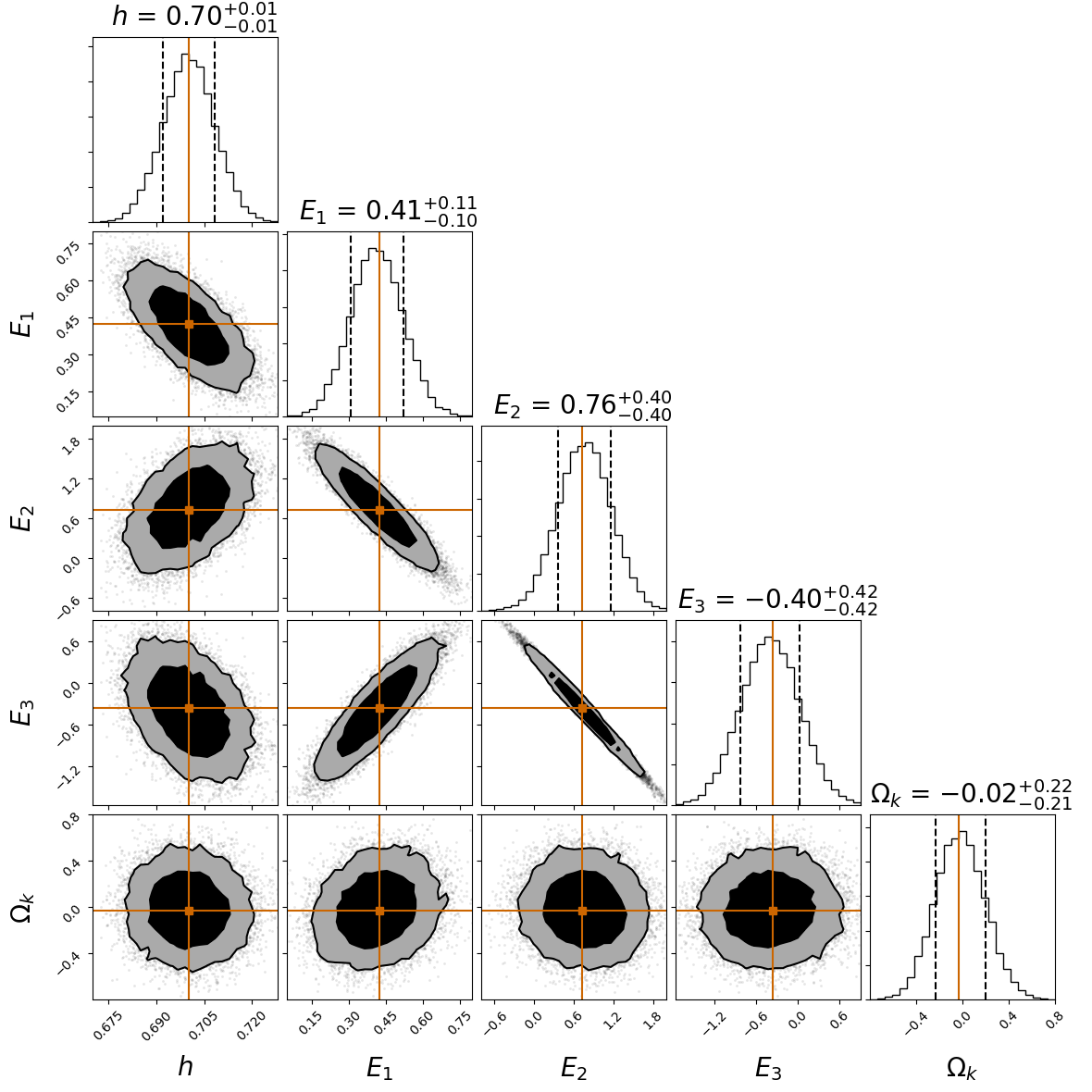} 
\end{center}
\caption{\small This figure shows the confidence contours on the 2D parameter spaces and the marginalised posterior distributions of the set of parameters $(h,E_1,E_2,E_3,\Omega_k)$, obtained in the combined analysis with OHD+SNe data. The corresponding 1-$\sigma$, 2-$\sigma$ contours are shown. In each panel, on top of the posterior probability distributions, the median  value of the parameters with associated $\pm$1-$\sigma$ errors are also quoted.}
\label{eispar}
\end{figure}
\begin{figure}[tb]
\begin{center}
\includegraphics[width = 0.95\textwidth]{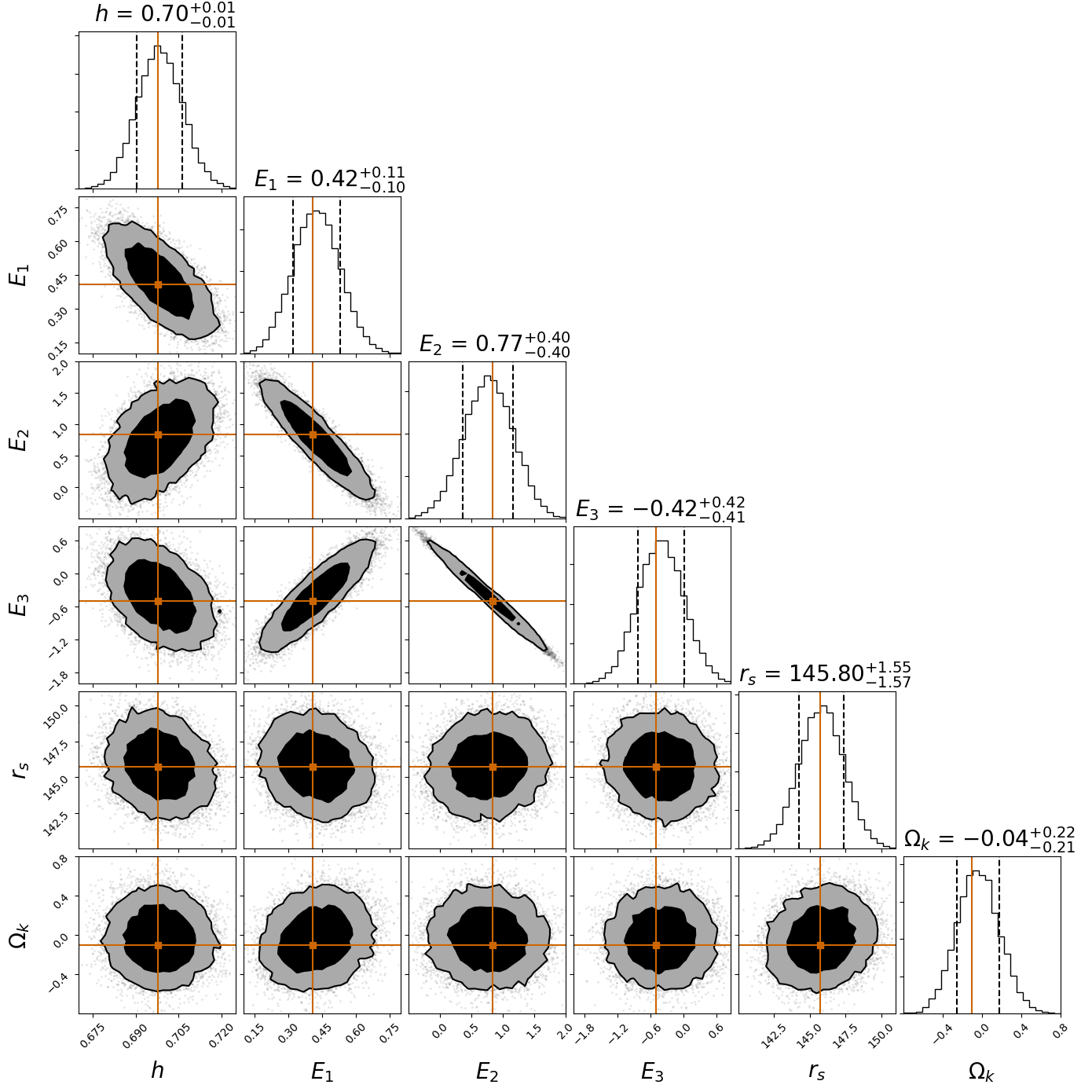} 
\end{center}
\caption{\small This figure shows the confidence contours on the 2D parameter spaces and the marginalised posterior distributions of the set of parameters $(h,E_1,E_2,E_3,r_s,\Omega_k)$, obtained in the combined analysis with OHD+SNe+BAO data. The corresponding 1-$\sigma$, 2-$\sigma$ contours are shown. In each panel, on top of the posterior probability distributions, the median  value of the parameters with associated $\pm$1-$\sigma$ errors are also quoted.}
\label{eispar_sne_ohd_bao}
\end{figure}
\begin{table}[t]
\caption{{\small This table shows the constraints on the present value of rescaled Hubble parameter $h$, $Eis$ parameters and the curvature contribution $\Omega_k$,  obtained from the analysis with two different combinations of the data sets. Constraint on acoustic sound horizon $r_s$ at photon drag epoch is obtained only in case of BAO data. The value estimated is $r_s=145.72(145.80^{+1.55}_{-1.57})$. The best-fit values (outside parentheses) and the median with associated $\pm$1-$\sigma$ uncertainties (inside parentheses) are quoted. The last column quotes the value of $\chi^2$ per degree of freedom as a measure of the goodness of fit. }}
\begin{center}
\resizebox{0.9990\textwidth}{!}{  
\begin{tabular}{c |c |c |c |c  |c  |c} 
  \hline
 \hline
   & $h$ & $E_1$ &  $E_2$ & $E_3$ & $\Omega_k$ & $\chi^2 / \rm dof$\\ 
 \hline
 OHD+SNe & $0.70~(0.70^{+0.01}_{-0.01})$ & $0.42~(0.41^{+0.11}_{-0.10})$ & $0.72~(0.76^{+0.40}_{-0.40})$ & $-0.36~(-0.40^{+0.42}_{-0.42})$ & $-0.03(-0.02^{+0.22}_{-0.21})$    & $49.64/62$\\ 
 \hline
 OHD+SNe+BAO & $0.70~(0.70^{+0.01}_{-0.01})$ & $0.41~(0.42^{+0.11}_{-0.10})$ & $0.84~(0.77^{+0.40}_{-0.40})$ & $-0.50~(-0.42^{+0.42}_{-0.41})$ & $-0.11(-0.04^{+0.22}_{-0.21})$    & $52.53/66$ \\ 
  \hline
\hline
\end{tabular}
}
\end{center}
\label{resulttable_e1e2}
\end{table}

\begin{figure}[tb]
\begin{center}
\includegraphics[angle=0, width=0.3\textwidth]{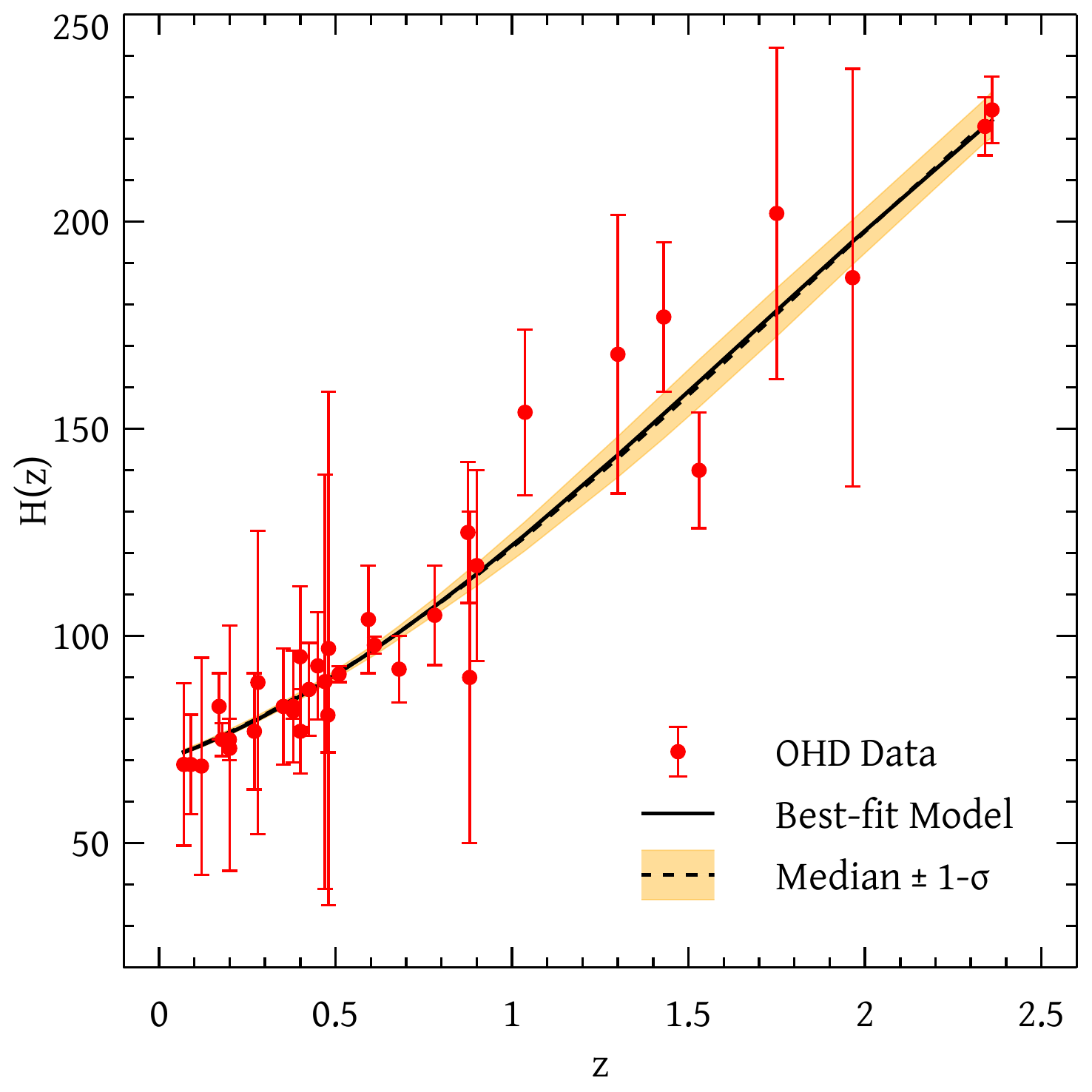}
\,\,
\includegraphics[angle=0, width=0.3\textwidth]{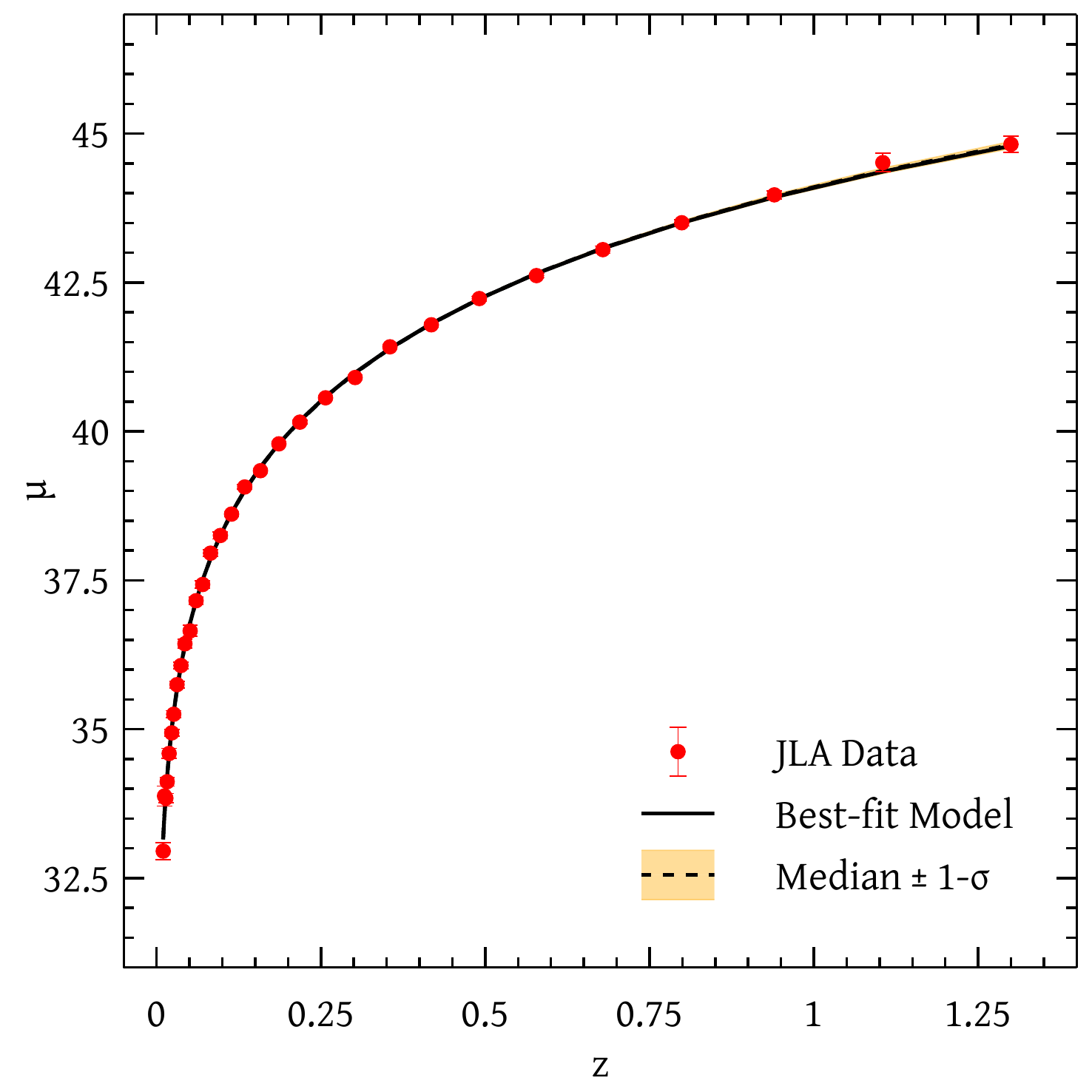}
\,\,
\includegraphics[angle=0, width=0.3\textwidth]{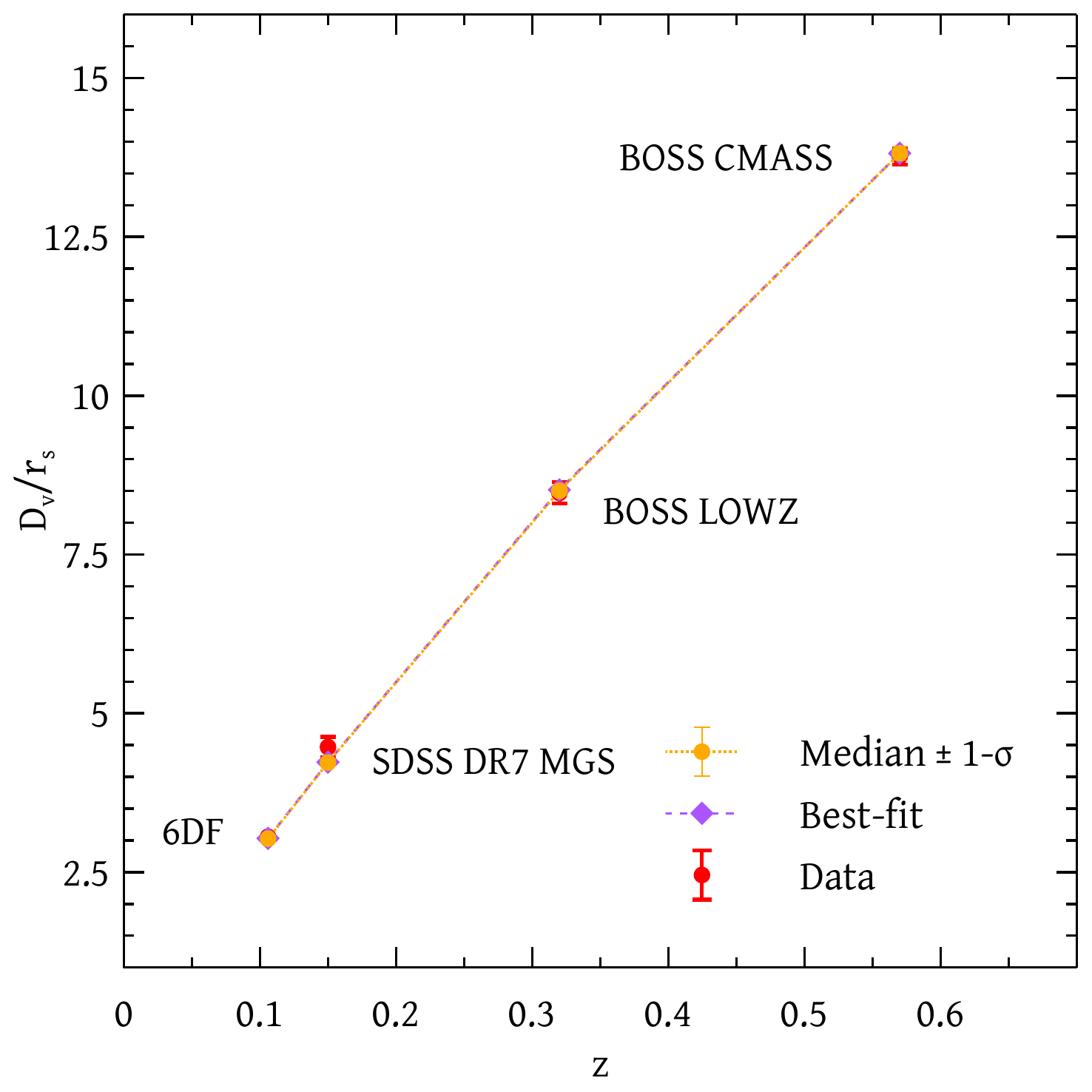}
\end{center}
\caption{\small This figure shows the observational data sets, used in the present analysis. Red points with error-bars denote observational measurements. The best-fit curves are shown as solid black lines. Dotted black curve with golden yellow band shows the median with $\pm$1-$\sigma$ confidence region.  Left panel shows the OHD data, middle one is for the JLA binned data points and the right panel shows the BAO data, namely the dilation scale ($D_v(z)$), scaled by acoustic sound horizon at photon drag epoch.}
\label{eisfit}
\end{figure}

Table \ref{resulttable_e1e2} presents the results obtained in the statistical analysis using different combinations of the data sets. The constraint on the present value of the rescaled Hubble parameter $h$, where $h=H_0/(100~\rm km~ sec^{-1}Mpc^{-1})$, and the coefficients of the Taylor expansion of the Hubble parameter $E_1$, $E_2$ and $E_3$ (from equation \ref{hubexp}) are shown. In the present analysis, we kept the curvature parameter $\Omega_k$ as a free parameter and constrain its value. Some recent analysis, by Yu, Ratra and Wang \cite{Yu:2017iju}, Ryan, Doshi and Ratra \cite{Ryan:2018aif}, Park and Ratra \cite{Park:2018tgj}, show that the contribution of spatial curvature is not so marginal as believed earlier. From the value of $\chi^2$, it is clear that our parametric model can describe the observational data sets quite well. Figure \ref{eispar} shows the confidence contours on different 2D parameter spaces and also the marginalised posterior distribution of the individual parameters, obtained from the combined analysis with OHD+SNe data. Figure \ref{eispar_sne_ohd_bao} shows the same obtained in the analysis with OHD+SNe+BAO. We took the comoving sound horizon ($r_s$) at photon drag epoch as a free parameter while using the BAO data. The value of $r_s$, obtained in the present analysis, is consistent with the recent non-parametric estimation by Haridasu {\it et al.} \cite{Haridasu:2018gqm}. We see that the uncertainty in the values of the parameters increases as we go towards higher order coefficients of the Taylor expansion of the Hubble parameter. The confidence contours on 2D parameter spaces show that $h$ has a negative correlation with $E_1$ and $E_3$  and a positive correlation with $E_2$. On the other hand, $E_1$ has negative correlation with $E_2$ and positive correlation with $E_3$, and consequently $E_2$ and $E_3$ are negatively correlated. Figure \ref{eisfit} shows the data points with error bars, used in the present analysis along with the best fit and the median curves with associated $\pm$1-$\sigma$ confidence regions obtained in the combined analysis with OHD+SNe+BAO data. The result, shown in table \ref{resulttable_e1e2} shows that the addition of BAO data does not change the result significantly. It is clear from the 2D contours, shown in figure \ref{eispar} and \ref{eispar_sne_ohd_bao}, that the correlations of curvature parameters $\Omega_k$ with the scaled Hubble constant ($h$) and $Eis$ parameters are very mild. The value of the scaled Hubble constant ($h$), obtained in the $Eis$ method, is consistent with Planck $\Lambda$CDM estimation \cite{Ade:2015xua} and the Dark Energy Survey result \cite{Macaulay:2018fxi} at 2-$\sigma$ level. The result is highly consistent with the recent model dependent estimation of $H_0$ by Park and Ratra \cite{Park:2018tgj} and also with the non-parametric estimation by Haridasu {\it et al.} \cite{Haridasu:2018gqm}.

In \cite{Aviles:2016wel}, the $Eis$ parameters are constrained using only the supernova distance modulus data, namely the JLA and Union 2.1 compilation. In the present analysis, we incorporated the OHD and BAO measurements also. The results, obtained in the present analysis, are consistent with the results of reference \cite{Aviles:2016wel}. The parameters have better convergence in the present case as the analysis has been carried out combining different data sets.

%
%
\section{Cosmographical parameters}
\label{Cosmographicalpar}
\begin{figure}[tb]
\begin{center}
\includegraphics[angle=0, width=0.95\textwidth]{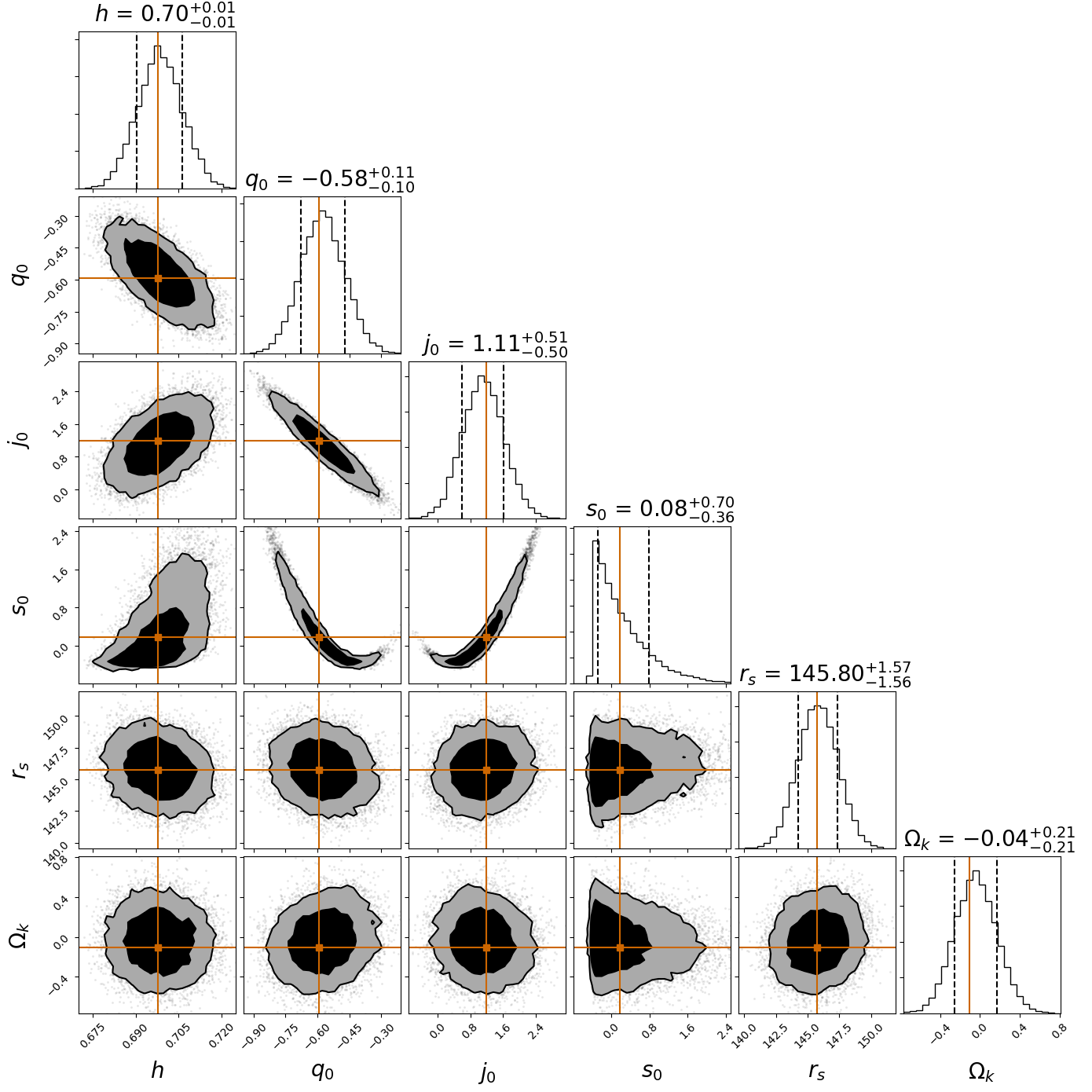}
\end{center}
\caption{\small This figure shows the confidence contours on the 2D parameter spaces and the marginalised posterior distributions of the set of parameters $(h,q_0,j_0,s_0,\Omega_k,r_s)$. The constraints on the cosmographical parameters are obtained from the values of {\it Eis} parameters estimated in the combined analysis with OHD+SNe+BAO data. The corresponding 1-$\sigma$ and 2-$\sigma$ contours are shown in the plot. The median value of the parameters with $\pm$1-$\sigma$ uncertainty is also quoted.}
\label{q0j0cont}
\end{figure}

Kinematical quantities are defined in terms of the scale factor and its time derivatives. The physically important kinematical quantities are the Hubble parameter ($H$), deceleration parameter ($q$), cosmological jerk parameter ($j$), snap parameter ($s$) and so on. Except for the Hubble parameter, others are dimensionless representations of different order time derivatives of the scale factor. The deceleration parameter, a dimensionless measurement of the acceleration of the universe, is defined as,  
\be
q=-\frac{1}{aH^2}\frac{d^2a}{dt^2}.
\ee
A negative value of the deceleration parameter indicates an accelerated expansion. The jerk parameter, the dimensionless representation of the third order time derivative of the scale factor, is defined as
\be
j=\frac{1}{aH^3}\frac{d^3a}{dt^3}.
\ee
\begin{table}[tb]
\caption{{\small Constraints on the cosmographical parameters obtained from the values of the {\it Eis} parameters in the present analysis. The best fit values (outside parentheses) and the median with 1$\sigma$ uncertainty (inside parentheses) are presented.}}
\begin{center}
\resizebox{0.8\textwidth}{!}{  
\begin{tabular}{c |c |c |c} 
  \hline
 \hline
   & $q_0$ &  $j_0$ & $s_0$\\ 
 \hline
 OHD+SNe & $-0.58(-0.59^{+0.11}_{-0.11})$ & $1.05(1.11^{+0.52}_{-0.51})$ & $0.04(0.10^{+0.74}_{-0.38})$\\ 
 \hline
 OHD+SNe+BAO & $-0.59(-0.58^{+0.11}_{-0.10})$ & $1.19(1.11^{+0.51}_{-0.50})$ & $0.18(0.08^{+0.70}_{-0.36})$\\ 
  \hline
  \hline
\end{tabular}
}
\end{center}
\label{resulttable_q0j0}
\end{table}
We can go further with higher order derivatives, like the snap parameter,
\be
s=\frac{1}{aH^4}\frac{d^4a}{dt^4},
\ee
and so on. The values of these kinematical quantities can be easily estimated if the distribution of different components in the matter sector is specified a priory. However, a model-independent estimation of these parameters is difficult due to the lack of suitable data. The evolution of the deceleration parameter is highly degenerate for different dark energy models. The higher order kinematic terms can potentially break this degeneracy. However, in a model-independent approach, the uncertainty will increase significantly as we go towards higher order kinematical terms. A semi model-dependent approach is more useful to constrain the kinematical quantities. One of the prescriptions in this direction is the {\it cosmography} \cite{weinberg,harrison,Bernstein:2003es,Visser:2004bf,Dabrowski:2004hx,Dabrowski:2005fg,Cattoen:2007sk}, where the luminosity distance is expressed in terms of the present values of the kinematical parameters as,
{\footnotesize
\bea
d_L(z) = \frac{c}{H_0}[z+\frac{1}{2}(1-q_0)z^2+\frac{1}{6}(-1+q_0+3q_0^2-j_0)z^3 
+\frac{1}{24}(2+5j_0-2q_0+10j_0q_0-15q_0^2-15q_0^3+s_0)z^4+...],
\label{cosgrap}
\eea
}
where the subscript index `0' indicates the present values of the corresponding parameters. These parameters ($q_0,j_0,s_0,...$) are often termed as {\it cosmographical parameters}.  Aviles, Klapp and Luongo \cite{Aviles:2016wel} have shown that the estimation of the parameters in this method suffers from the problem of bias and this is only suitable for those observables, which are directly related to the luminosity distance, for instance, the distance modulus measurements of type Ia supernovae. However, it is not very useful for other observational data sets. To overcome these issues, they prescribed a Taylor series expansion of the Hubble parameter (equation \ref{hubexp}), namely the {\it Eis} method, as a better methodology to estimate the kinematic parameters in a model independent way.  The relations between the {\it Eis} parameters and the cosmographical parameters are,
\begin{align}
q_0 &= -1 + E_1  \,\, , \label{qeis}\\ 
j_0 &= q_0^2 + E_2 \,\, , \label{jeis}\\
s_0 &=3 ( q_0^2 + q_0^3 ) - j_0 (4 q_0 + 3) - E3 \,\,. \label{seis} 
\end{align}
In the present analysis, the values of the cosmographical parameters have been obtained from the estimated values of the Eis parameters. Table \ref{resulttable_q0j0} presents the results, the estimated values of the parameters and the associated $\pm$1-$\sigma$ uncertainty. It is important to note that the uncertainties associated with the higher order kinematical terms are larger. Figure \ref{q0j0cont} shows the confidence contours on different 2D parameter spaces and the marginalised posterior distribution of the cosmographical parameters. The scaled Hubble constant ($h$) is negatively correlated with $q_0$ and has a positive correlation with $j_0$ and $s_0$. On the other hand, $q_0$ has a negative correlation with $j_0$ and $s_0$. Consequently, $j_0$ and $s_0$  are positively correlated. The curvature contribution $\Omega_k$ has very weak correlations with the cosmographical parameters.

\begin{figure}[tb]
\begin{center}
\includegraphics[angle=0, width=0.3\textwidth]{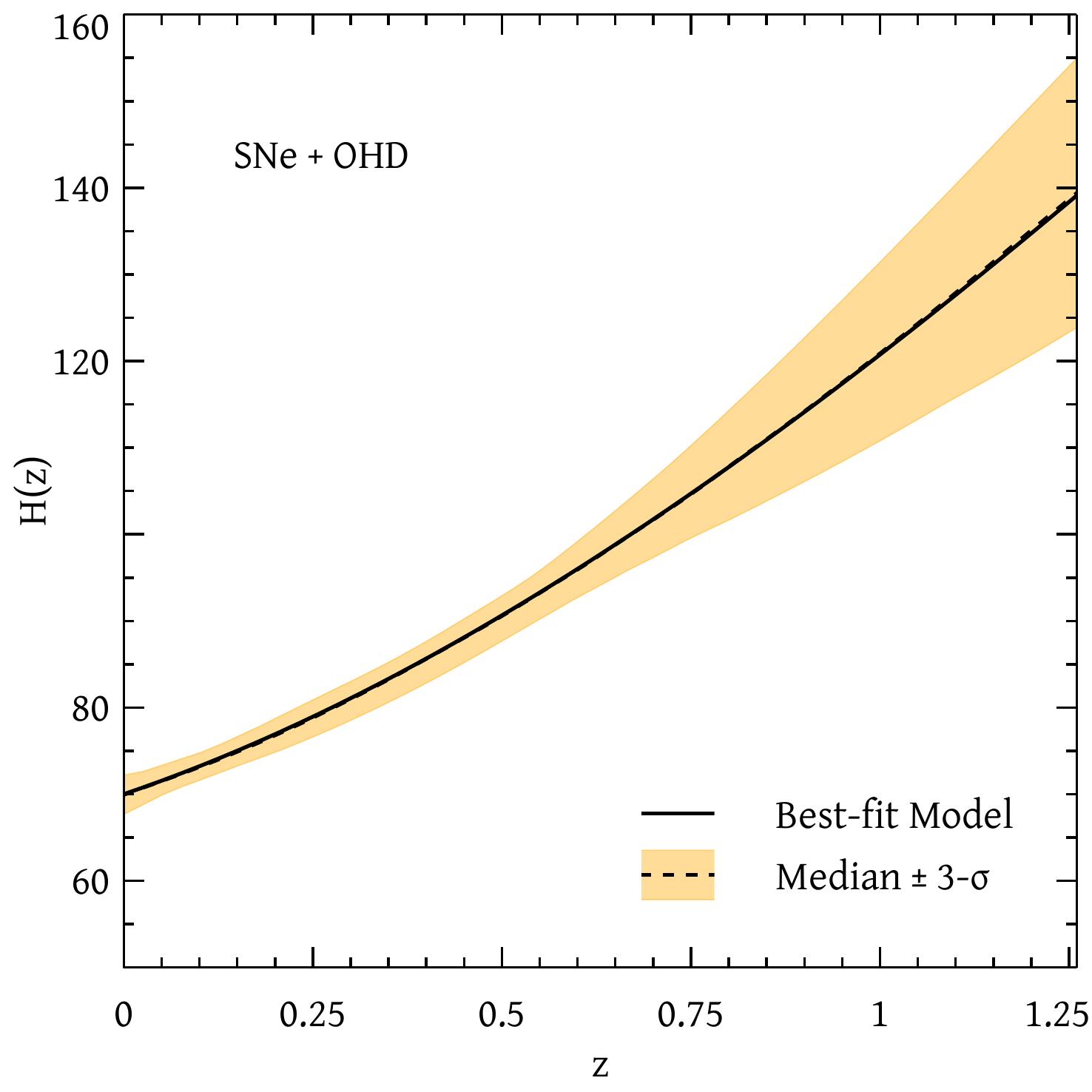}
\,\,
\includegraphics[angle=0, width=0.3\textwidth]{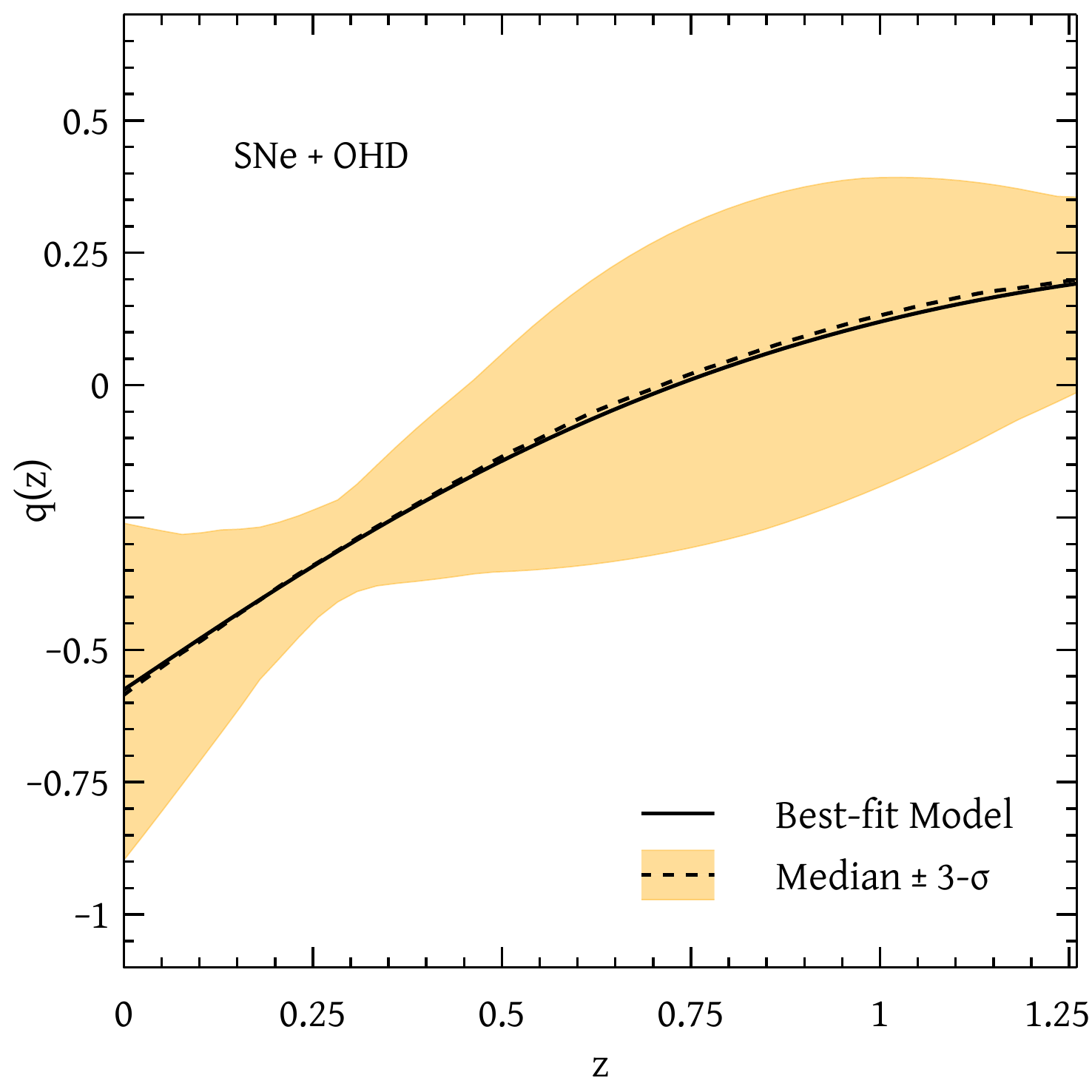}
\,\,
\includegraphics[angle=0, width=0.3\textwidth]{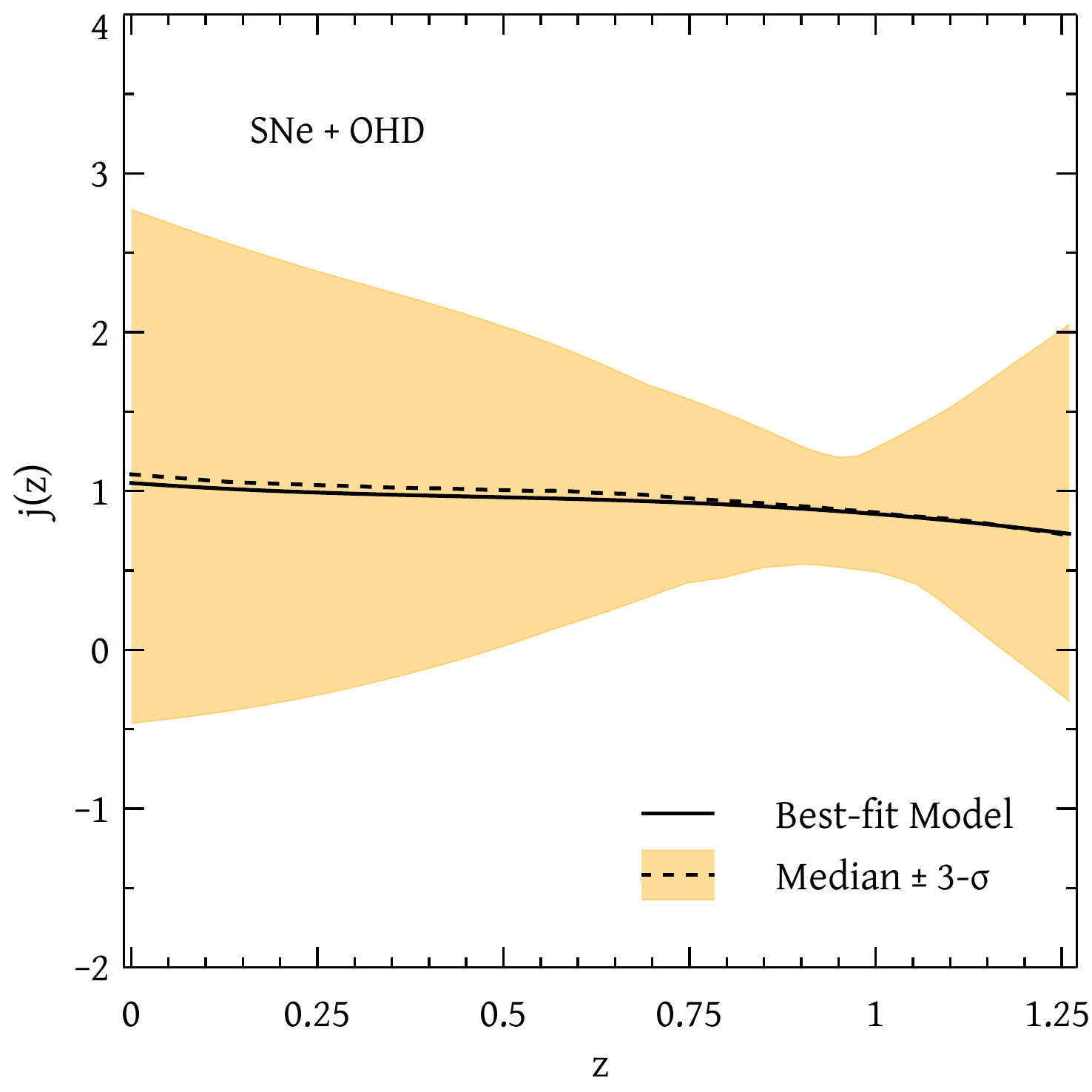}
\,\,
\includegraphics[angle=0, width=0.3\textwidth]{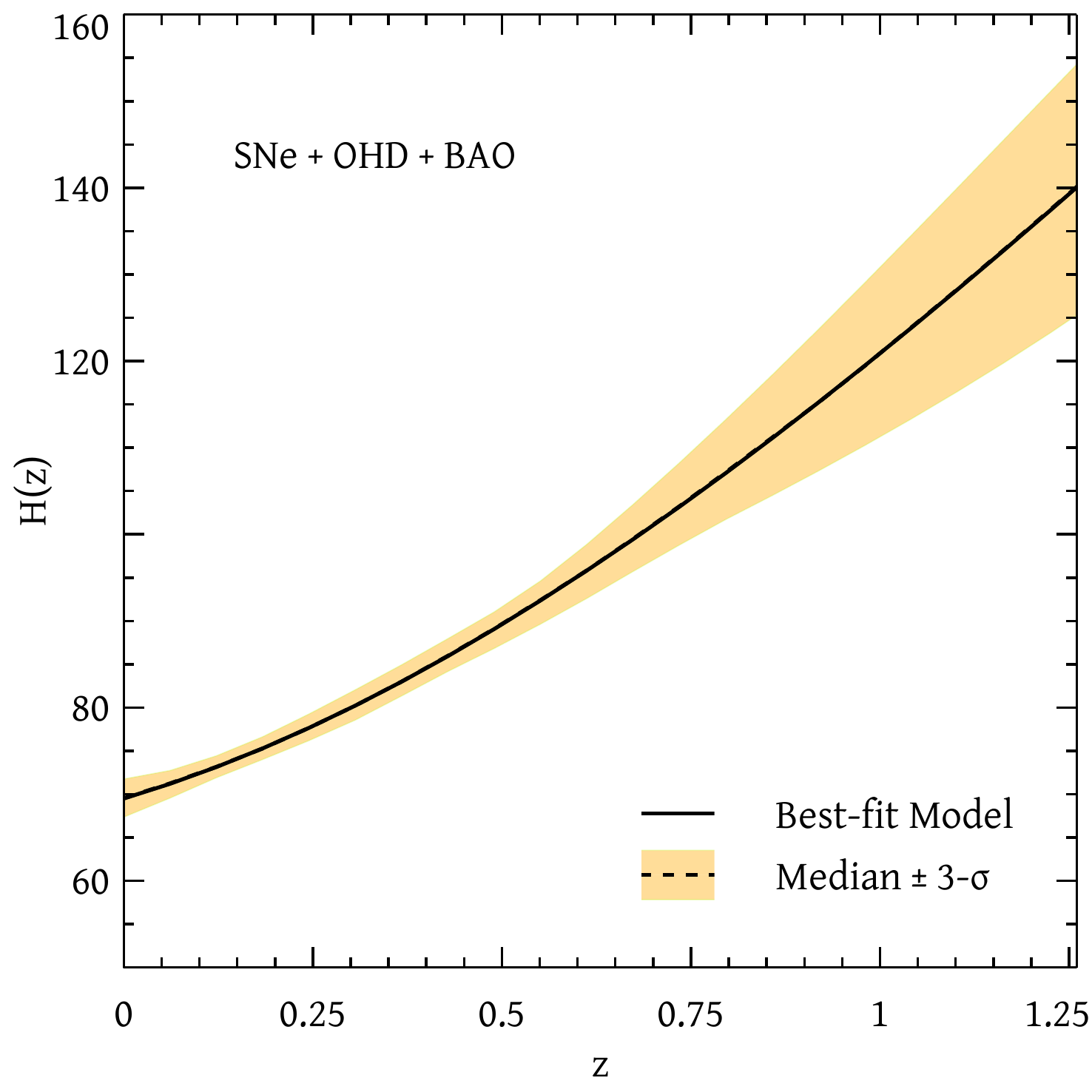}
\,\,
\includegraphics[angle=0, width=0.3\textwidth]{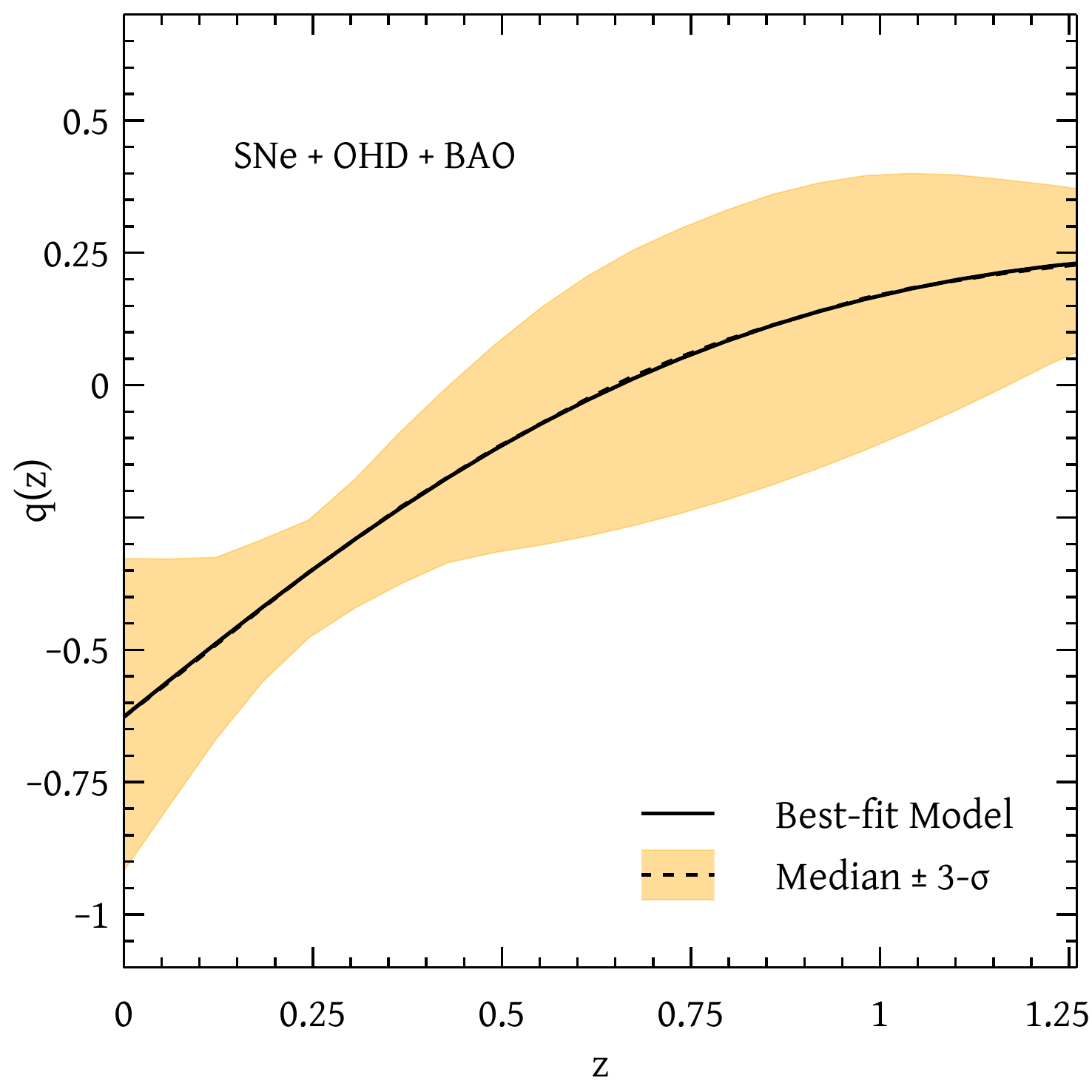}
\,\,
\includegraphics[angle=0, width=0.3\textwidth]{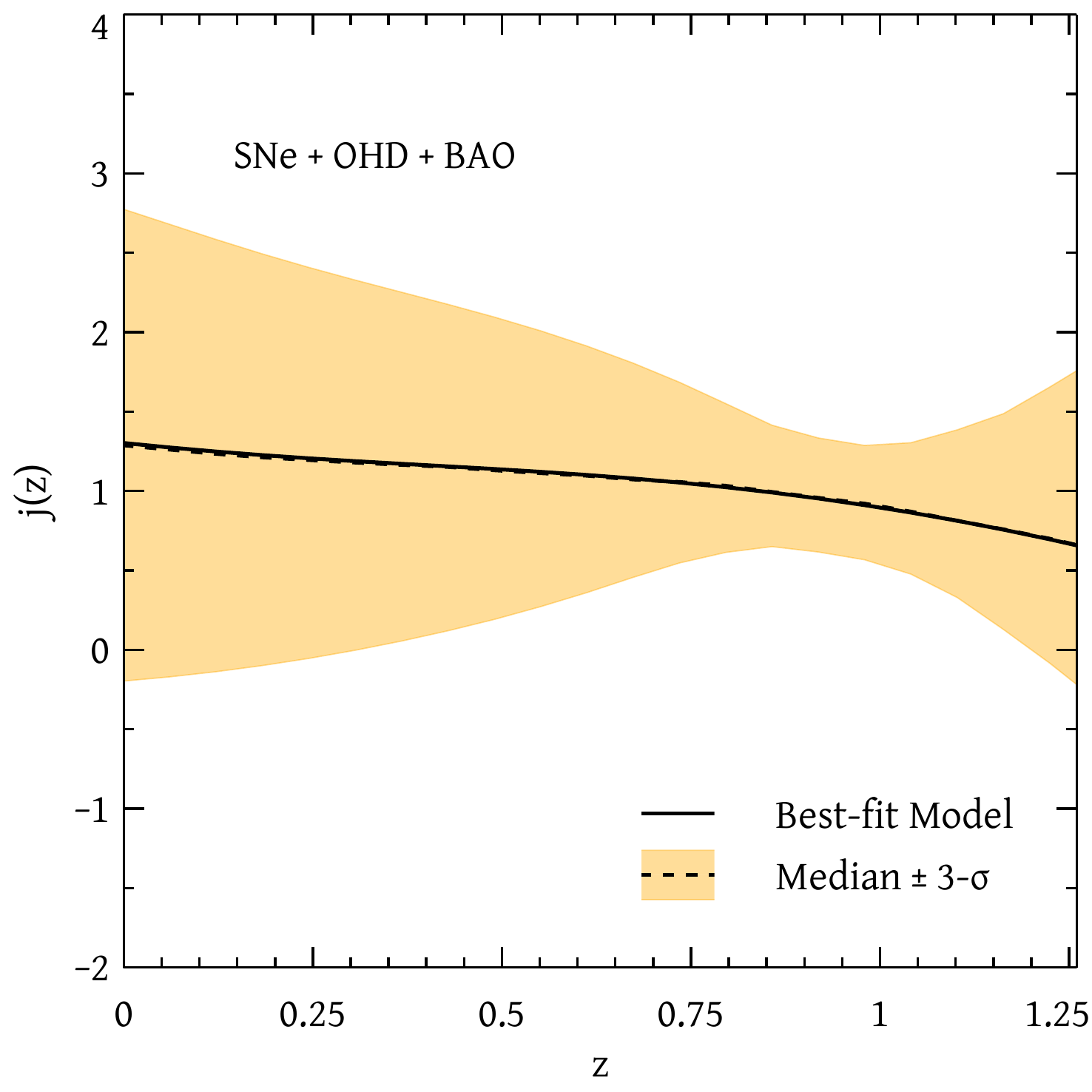}
\end{center}
\caption{\small The evolution of H(z), q(z) and j(z) for two different combinations of the data sets, namely OHD+SNe (upper panels) and OHD+SNe+BAO (lower panels). The best fit curve and the median along with the uncertainty extended up to 3-$\sigma$ confidence regions are shown.}
\label{Hqj_plots}
\end{figure} 
Figure \ref{Hqj_plots} shows the evolution of different kinematical parameters, namely $H(z)$, $q(z)$ and $j(z)$. The best fit and the median with associated 3-$\sigma$ confidence region, obtained in the {\it Eis} method, are shown. The plots in figure \ref{Hqj_plots} clearly reveal that $H(z)$ and  $q(z)$ are well constrained from the present observations, but the uncertainty increases enormously in $j(z)$. The plots also reveal that the addition of BAO data does not bring any significant improvement in the parameter constraints in the present analysis.
%
%
\section{Constraints on the $Statefinder$ and $Om$ diagnostic}
\label{stateOm}
\begin{figure}[tb]
\begin{center}
\includegraphics[angle=0, width=0.95\textwidth]{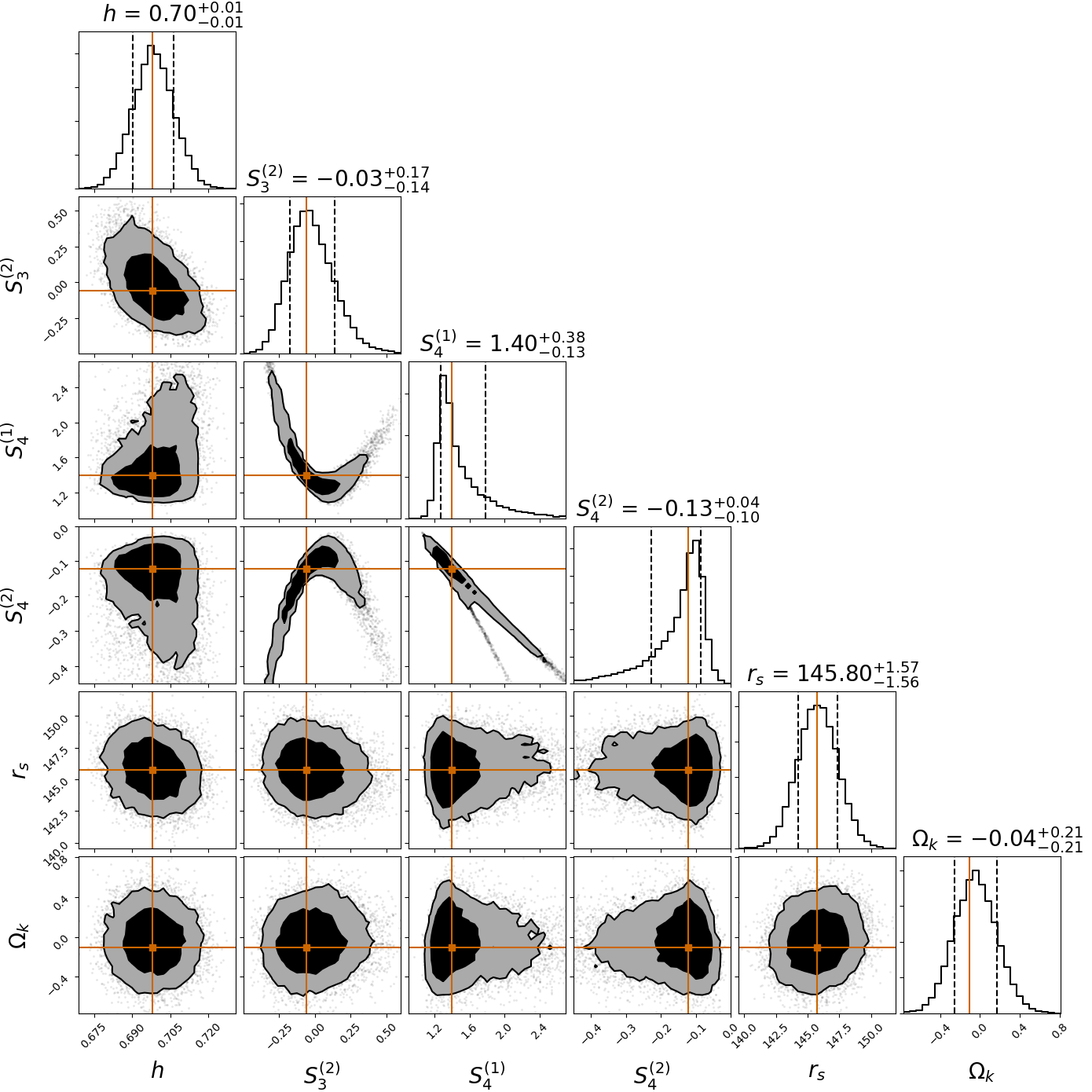}
\end{center}
\caption{\small This figure shows the confidence contours on the 2D parameter spaces and the marginalised posterior distributions of the present values of the statefinder parameters $(S_3^{(2)},S_4^{(1)},S_4^{(2)})$ along with the scaled Hubble constant $(h)$, curvature contribution $\Omega_k$ and acoustic sound horizon $r_s$, obtained in the combined analysis with OHD+SNe+BAO data. The corresponding 1-$\sigma$ and 2-$\sigma$ contours are shown. The median value of the parameters with $\pm$1-$\sigma$ uncertainty is also quoted.}
\label{S3S4cont}
\end{figure}

The  {\it Statefinder} parameters, which are defined in terms of the cosmological expansion factor and its derivatives, are used as a diagnostic of {\it concordance cosmology} ($\Lambda$CDM). In a recent work, Arabsalmani and Sahni \cite{Arabsalmani:2011fz} have introduced  the {\it statefinder hierarchy} which contains the higher order time derivatives of the scale factor, i.e. $\frac{d^na}{dt^n}$ , $n\geq 2$. The Taylor expansion of the scale factor is given as, 

\be 
(1+z)^{-1}=\frac{a(t)}{a_0}=1+\sum_{n=1}^{\infty}\frac{A_n(t_0)}{n!}H^n_0(t-t_0)^n,
\label{taylor_a} 
\ee
where $A_n=\frac{a^{(n)}}{aH^n}$. The $a^{(n)}$ is the $n^{th}$ order time derivative of the scale factor and $t_0$ is the present time. The way the coefficients $A_n$ are defined, it is clear that $A_2=-q$, $A_3=j$ and $A_4=s$. In $\Lambda$CDM cosmology, all the $A_n$-s can be written in terms of the matter density parameter ($\Omega_{m0}$) which is the present matter density (dark matter+baryonic matter) scaled by the present critical density of the universe $(3H_0^2/8\pi G)$. The relations are given as,
\be
-q:=A_2=1-\frac{3}{2}\tilde{\Omega}_m,
\label{a2}
\ee
\be
j:=A_3=1,
\label{a3}
\ee
\be
s:=A_4=1-\frac{3^2}{2}\tilde{\Omega}_m.
\label{a4}
\ee
where $\tilde{\Omega}_m=\Omega_{m0}(1+z)^3/E^2(z)$. From equation (\ref{a2}), $\tilde{\Omega}_m$ can be written in terms of $q$. Thus from equation (\ref{a4}), we get,
\be
s+3(1+q)=1.
\label{a5}
\ee 
Its is important to mention it again that equations (\ref{a2}) to (\ref{a5}) are valid only for $\Lambda$CDM cosmology. Now we can define the statefinder  parameters  as 
\be
S_3^{(1)}=A_3,
\label{s3}
\ee
\be
S_4^{(1)}=A_4+3(1+q).
\label{s4}
\ee
In a similar way, higher order statefinders can be defined. In the present work, we are restricted only upto $S_4^{(1)}$ as the observational constraint obtained up to the fourth order time derivative of the scale factor in the present analysis. This set of statefinders parameters have the value of unity for $\Lambda$CDM. Now, another set of statefinders can be defined from the first set ($S_n^{(1)}$) as,
\be
S_n^{(2)}=\frac{S_n^{(1)}-1}{\alpha(q-\frac{1}{2})},
\label{sn2}
\ee
where $\alpha$ is an arbitrary constant. The second set of statefinders is called {\it fractional statefinder} \cite{Arabsalmani:2011fz}. In $\Lambda$CDM cosmology, the second statefinders $S_n^{(2)}=0$ and thus a set of null diagnostic of dark energy can be defined as, 
\be
\{S_n^{1}, S_n^{(2)}\}=\{1,0\}.
\ee
Depending upon the value of $n$, it forms a hierarchy of the null diagnostics, stated as the {\it statefinder hierarchy}. The statefinder parameters at redshift $z=0$ are directly connected to the cosmographical parameters $(q_0,j_0,s_0)$. Thus we can constrain the statefinder hierarchy at redshift $z=0$. The present value of the statefinder parameters, presented in the hierarchy, are connected to the cosmographical parameters ($q_0,j_0,s_0$) as,
\begin{align}
S_3^{(1)} &= j_0 \,\, , \\
S_3^{(2)} &= \frac{S_3^{(1)}-1}{3(q_0 - 1/2)} \,\, ,\\
S_4^{(1)} &= s_0 + 3(1+q_0) \,\, ,\\
S_4^{(2)} &= \frac{S_4^{(1)}-1}{3 (q_0 - 1/2)}.
\end{align}
The cosmographical parameters are directly related to the {\it Eis} parameters (equation (\ref{qeis}) to (\ref{seis})). Thus the present values of the sets of statefinder parameters are directly related to the {\it Eis} parameters. We used the estimated values of the {\it Eis} parameters to constrain the statefinders hierarchy. The results, presented in table \ref{resulttable_S3S4}, show that the corresponding $\Lambda$CDM values for the first set of statefinder hierarchy $\{S_3^{(1)}, S_3^{(2)}\}$ are well within the 1-$\sigma$ confidence region of the values estimated in the present kinematic approach. But for the second set of statefinder hierarchy, $\{S_4^{(1)}, S_4^{(2)}\}$, the corresponding $\Lambda$CDM value is not within the 1-$\sigma$ confidence region (table \ref{resulttable_S3S4}). Figure \ref{S3S4cont} shows the confidence contours and the posterior probability distribution function of the statefinder parameters alond with the scaled Hubble constant $h$, curvature contribution $\Omega_k$ and acoustic sound horizon $r_s$. As usual, the $\Omega_k$ and $r_s$ are found to be almost uncorrelated with the statefinder parameters. The Hubble constant also has a week correlation with the statefinder parameters. The nature of correlations of $S_4^{(1)}$ and $S_4^{(2)}$ with $S_3^{(2)}$ vary depending upon the value of $S_3^{(2)}$. On the other hand,  $S_4^{(1)}$ and $S_4^{(2)}$ are negatively correlated.
\begin{table}[tb]
\caption{{\small Constraints on the present values of the parameters of statefinder hierarchy obtained from the values of the {\it Eis} parameters in the present analysis. The best fit values (outside parentheses) and the median with $\pm$1-$\sigma$ uncertainty (inside parentheses) are presented.}}
\begin{center}
\resizebox{0.92\textwidth}{!}{  
\begin{tabular}{c |c |c } 
  \hline
 \hline
   & OHD+SNe & OHD+SNe+BAO  \\ 
 \hline
 $\{S_3^{(1)}, S_3^{(2)}\}$ & $\{1.05~(1.11^{+0.52}_{-0.51}), -0.02~(-0.03^{+0.17}_{-0.14})\}$ & $\{1.19~(1.11^{+0.51}_{-0.50}),-0.06~(-0.03^{+0.17}_{-0.14})\}$\\ 

\hline

 $\{S_4^{(1)}, S_4^{(2)}\}$ & $\{1.32~(1.39^{+0.41}_{-0.14}), -0.10~(-0.12^{+0.04}_{-0.11})\}$ & $\{1.40~(1.40^{+0.38}_{-0.13}),-0.12~(-0.13^{+0.04}_{-0.10})\}$\\ 
  \hline
  \hline
\end{tabular}
}
\end{center}
\label{resulttable_S3S4}
\end{table}

Another null diagnostic of dark energy is $Om(z)$, introduced by Sahni et al. \cite{Sahni:2008xx} and by Zunckel and Clarkson \cite{Zunckel:2008ti}. It is defined as,

\be
Om(z)=\frac{h^2(z)-1}{(1+z)^3-1}
\label{Om}
\ee

\begin{figure}[tb]
\begin{center}
\includegraphics[angle=0, width=0.4\textwidth]{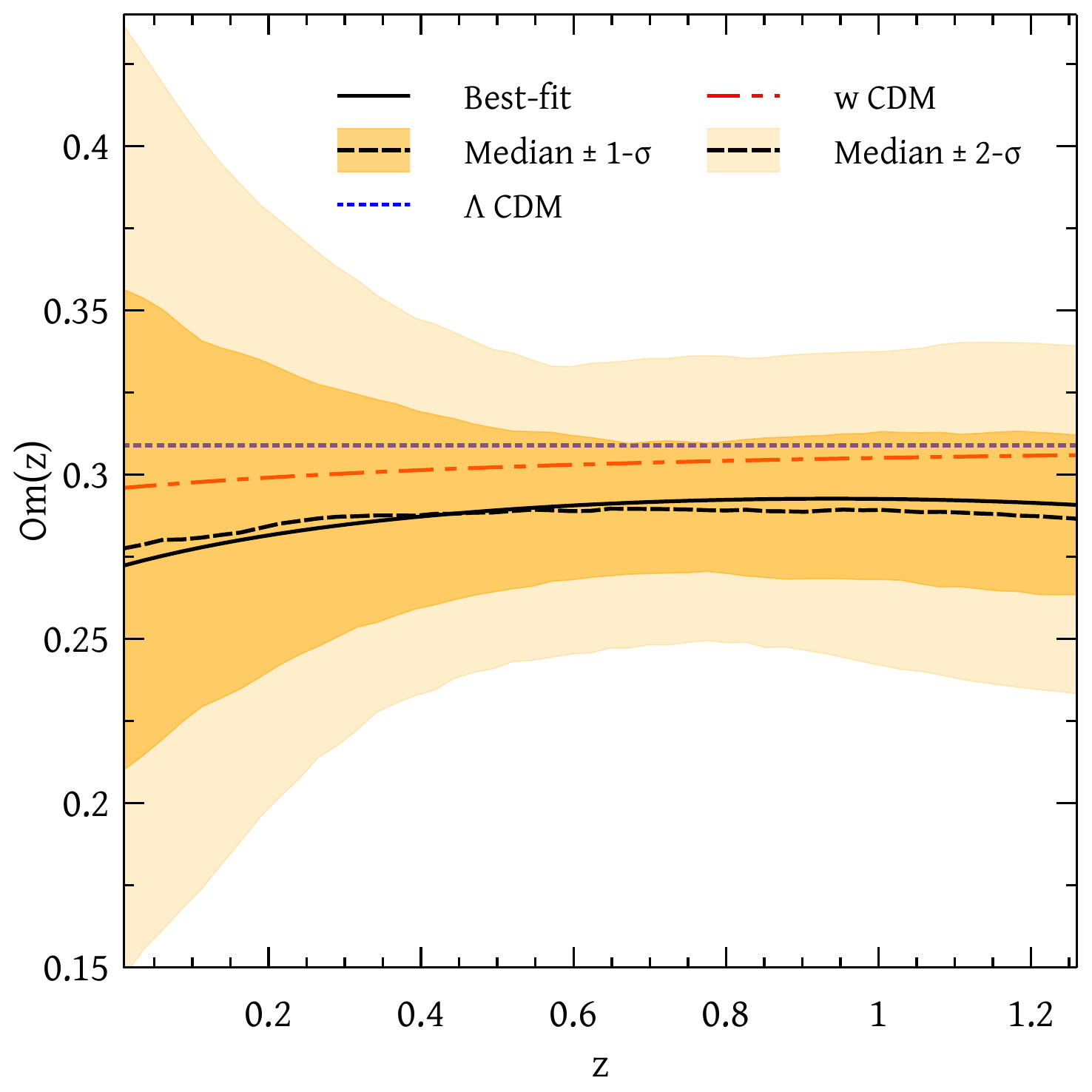}
\includegraphics[angle=0, width=0.4\textwidth]{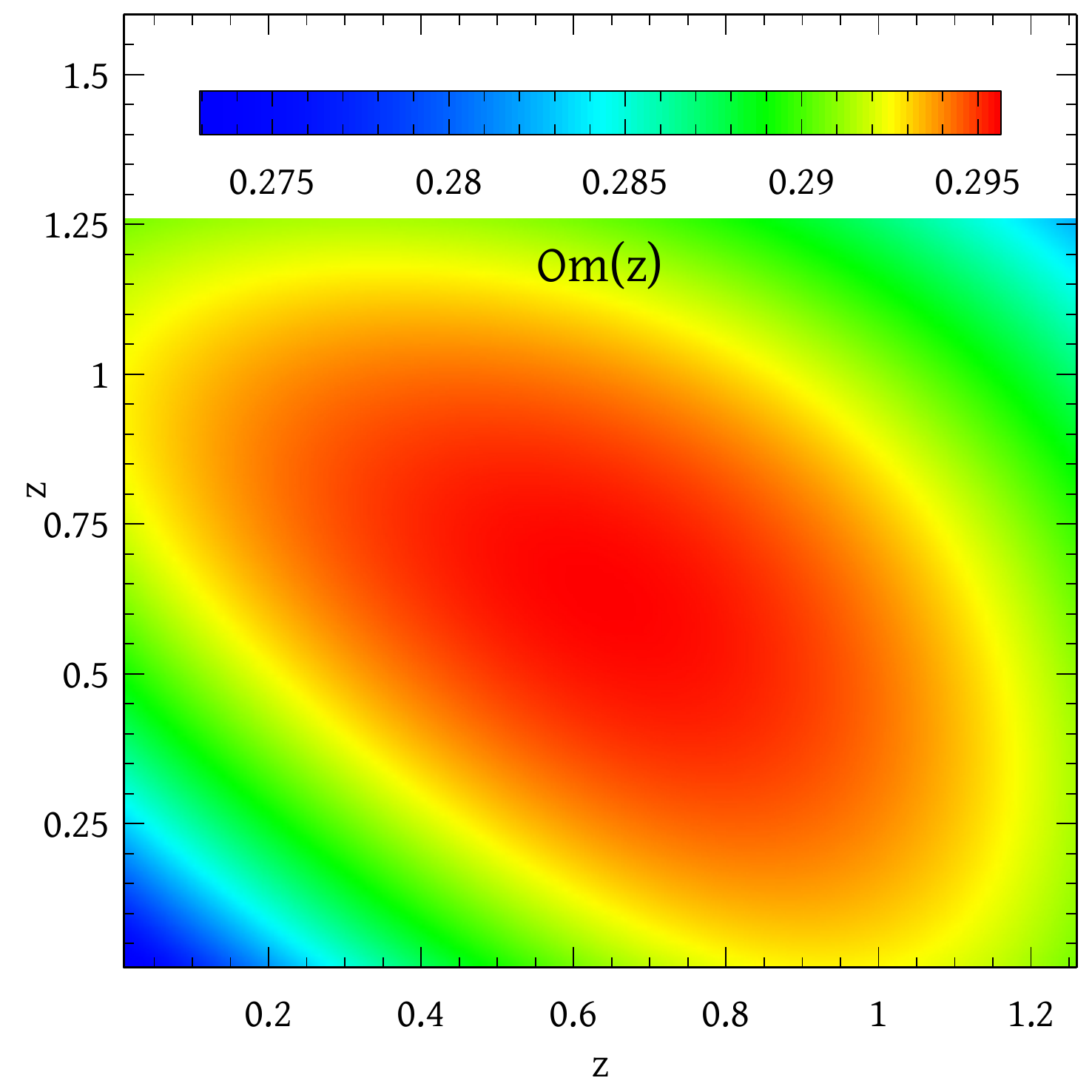}
\end{center}
\caption{\small Left panel of this figure shows  the $Om(z)$ plot (best fit and median with 1-$\sigma$ and 2-$\sigma$ confidence regions) for the present estimated values of the coefficients of the Taylor expansion of  the Hubble parameter. The right panel presents the two point function $Om(z_i;z_j)$; the variation in the value is shown by the variation of colour.}
\label{Om_plots}
\end{figure}

For $\Lambda$CDM cosmology, the value of $Om(z)$ remains constant which is equal to $\Omega_{m0}$, the present value of the matter density scaled by the present critical density. Evolution in the value of $Om(z)$ diagnostic carries the signature of deviation from $\Lambda$CDM. As the $Om$ only depends upon the expansion rate, it is easier to be determined from the present observations. The $Om$-diagnostics can also be defined as a two-point function \cite{Shafieloo2012},

\be
Om(z_i;z_j)=\frac{h^2(z_i)-h^2(z_j)}{(1+z_i)^3-(1+z_j)^3}
\label{Om2}
\ee
Figure \ref{Om_plots} presents the plot of $Om(z)$ (left panel) corresponding to the estimated values of the {\it Eis} parameters in the combined analysis with OHD+SNe+BAO. The two-point function $Om(z_i; z_j)$ is shown in the right panel of figure \ref{Om_plots} where the value of the function is represented by different colours. Both the $Om(z)$ and $Om(z_i; z_j)$ plots show a possibility of a slight variation of these quantities. The corresponding $\Lambda$CDM and $w$CDM values of $Om(z)$, obtained from Planck \cite{Ade:2015xua} estimation, are consistent with the reconstructed $Om(z)$ and $Om(z_i; z_j)$ at 1-$\sigma$ confidence level. 

%
%
\section{Conclusion}
\label{conclusion}
A kinematic approach to the reconstruction of a late-time cosmic acceleration model is independent of any prior assumption about the gravity theory or the distribution in the matter sector of the universe. The present work, where the statefinder parameters are constrained from the Taylor series expansion of the Hubble parameter, is independent of any assumption about the nature and evolution of dark energy and dark matter. The values of the cosmographical parameters and the set of dark energy diagnostics, obtained in the present analysis, can be compared with the values corresponding to different dark energy models. Thus the present analysis is useful for selecting a viable dark energy scenario. The statefinder parameters can be represented as a set of null diagnostics of $\Lambda$CDM model, as discussed by Arabsalmani and Sahni \cite{Arabsalmani:2011fz}. The values of the parameters, presented in the hierarchy of statefinder diagnostic, are also constrained through the present kinematic approach. The values of the first set of parameters in the statefinder hierarchy allow the corresponding $\Lambda$CDM value within 1-$\sigma$ confidence region. But the second set of parameters in the hierarchy ($\{S_4^{(1)},S_4^{(2)}\}$) shows a disagreement with the corresponding $\Lambda$CDM values at the 2-$\sigma$ level. However, at this point, this is not sufficient to conclude this as a serious tension of concordance cosmology. The $Om(z)$ and $Om(z_i,z_j)$ diagnostics are showing good agreement to the concordance cosmology. The allowed values of the parameters in the statefinder hierarchy can be used as a probe towards the viability of different dark energy models. At this point, it is important to mention that the present analysis has its own limitation as it has been carried out only with OHD, SNe and BAO data. A detailed analysis incorporating the CMB likelihood might have its signature on the result. However, the CMB likelihood cannot be easily used in the analysis as the evolution of different components in the energy budget are not known in case of a model-independent reconstruction.  

\par Though the present analysis is a parametric approach, this can alleviate the difficulty of a purely model-independent non-parametric approach to some extent. Better constraints on the higher-order kinematical quantities and the hierarchy of the set of statefinder diagnostics are obtained in the present approach which is hardly possible with a purely model-independent statistical method with the currently available observational data sets.
%
%
\section*{Acknowledgment}
We sincerely acknowledge the use of high performance computing facilities at IUCAA, Pune. This work used the open source computing packages NumPy \cite{numpy}, SciPy \cite{scipy} and the plotting softwares Veusz \cite{veusz} and corner.py \cite{Foreman-Mackey2016}. NP acknowledges the financial support from the Council of Scientific and Industrial Research (CSIR), India as a Shyama Prasad Mukherjee Fellow. Finally, we would like to acknowledge the anonymous referee for valuables comments and suggestions that led to a substantial improvement of the work.


\begin{thebibliography}{}


\bibitem{snia1} 
  A.~G.~Riess {\it et al.} [Supernova Search Team],
  Astron.\ J.\  {\bf 116}, 1009 (1998).

\bibitem{snia2} 
  S.~Perlmutter {\it et al.} [Supernova Cosmology Project Collaboration],
  Astrophys.\ J.\  {\bf 517}, 565 (1999).



\bibitem{Riess:2004nr} 
  A.~G.~Riess {\it et al.} [Supernova Search Team],
  Astrophys.\ J.\  {\bf 607}, 665 (2004).




\bibitem{Carroll:2000fy} 
  S.~M.~Carroll,
  Living Rev.\ Rel.\  {\bf 4}, 1 (2001).

\bibitem{Padmanabhan:2002ji} 
  T.~Padmanabhan,
  Phys.\ Rept.\  {\bf 380}, 235 (2003).




\bibitem{Sahni:1999gb} 
  V.~Sahni and A.~A.~Starobinsky,
  Int.\ J.\ Mod.\ Phys.\ D {\bf 9}, 373 (2000).

\bibitem{Sahni:2006pa} 
  V.~Sahni and A.~Starobinsky,
  Int.\ J.\ Mod.\ Phys.\ D {\bf 15}, 2105 (2006).


\bibitem{Peebles:2002gy} 
  P.~J.~E.~Peebles and B.~Ratra,
  Rev.\ Mod.\ Phys.\  {\bf 75}, 559 (2003).

\bibitem{Copeland:2006wr} 
  E.~J.~Copeland, M.~Sami and S.~Tsujikawa,
  Int.\ J.\ Mod.\ Phys.\ D {\bf 15}, 1753 (2006).





\bibitem{Chen:2016eyp} 
  Y.~Chen, B.~Ratra, M.~Biesiada, S.~Li and Z.~H.~Zhu,
  Astrophys.\ J.\  {\bf 829}, 61 (2016).


\bibitem{Park:2018fxx} 
  C.~G.~Park and B.~Ratra,
  arXiv:1807.07421 [astro-ph.CO].


\bibitem{Durrive:2018quo} 
  J.~B.~Durrive, J.~Ooba, K.~Ichiki and N.~Sugiyama,
  Phys.\ Rev.\ D {\bf 97}, 043503 (2018).

\bibitem{Ryan:2018aif} 
  J.~Ryan, S.~Doshi and B.~Ratra,
  Mon.\ Not.\ Roy.\ Astron.\ Soc.\  {\bf 480}, no. 1, 759 (2018).











\bibitem{Crittenden:2005wj} 
  R.~G.~Crittenden, L.~Pogosian and G.~B.~Zhao,
  JCAP {\bf 0912}, 025 (2009).



\bibitem{Clarkson:2010bm} 
  C. Clarkson and C. Zunckel,
  Phys.\ Rev.\ Lett.\  {\bf 104}, 211301 (2010).








\bibitem{Ishida:pca}
 E. E. O. Ishida and R. S. de Souza, 
 Astron. Astrophys.  {\bf 527}, A49 (2011).



\bibitem{Amendola:pca}
L. Amendola, A. C. O. Leite, C. J. A. P. Martins, N. J. Nunes, P. O. J. Pedrosa and A. Seganti,
Phys.\ Rev.\ D {\bf 86}, 063515 (2012).




\bibitem{Qin:2015eda} 
  H.~F.~Qin, X.~B.~Li, H.~Y.~Wan and T.~J.~Zhang,
  arXiv:1501.02971 [astro-ph.CO].







\bibitem{Holsclaw:gp}
T. Holsclaw, U. Alam, B. Sansó, H. Lee, K. Heitmann, S. Habib and D. Higdon,
 Phys.\ Rev.\ Lett.\ {\bf 105}, 241302 (2010);
 Phys.\ Rev.\ D {\bf 82}, 103502 (2010);
 Phys.\ Rev.\ D {\bf 84}, 083501 (2011).



\bibitem{Seikel:2012uu} 
  M.~Seikel, C.~Clarkson and M.~Smith,
  JCAP {\bf 1206}, 036 (2012).



\bibitem{Nair:2013sna} 
  R.~Nair, S.~Jhingan and D.~Jain,
  JCAP {\bf 1401}, 005 (2014).



\bibitem{Shafieloo:2012ht} 
  A.~Shafieloo, A.~G.~Kim and E.~V.~Linder,
  Phys.\ Rev.\ D {\bf 85}, 123530 (2012).





\bibitem{weinberg}S. Weinberg, {\it Gravitation and Cosmology: Principles and
Applications of the General Theory of Relativity}. (Wiley, New York, 1972).


\bibitem{harrison}E. R. Harrison, Nature (London) {\bf 260}, 591 (1976).


\bibitem{Bernstein:2003es} 
  G.~M.~Bernstein and B.~Jain,
  Astrophys.\ J.\  {\bf 600}, 17 (2004).



\bibitem{Visser:2004bf} 
  M.~Visser,
  Gen.\ Rel.\ Grav.\  {\bf 37}, 1541 (2005).




\bibitem{Dabrowski:2004hx} 
  M.~P.~Dabrowski and T.~Stachowiak,
  Annals Phys.\  {\bf 321}, 771 (2006).





\bibitem{Dabrowski:2005fg} 
  M.~P.~Dabrowski,
  Phys.\ Lett.\ B {\bf 625}, 184 (2005).





\bibitem{Cattoen:2007sk} 
  C.~Cattoen and M.~Visser,
  Class.\ Quant.\ Grav.\  {\bf 24}, 5985 (2007).








\bibitem{Aviles:2012ay} 
  A.~Aviles, C.~Gruber, O.~Luongo and H.~Quevedo,
  Phys.\ Rev.\ D {\bf 86}, 123516 (2012).

\bibitem{Dunsby:2015ers} 
  P.~K.~S.~Dunsby and O.~Luongo,
  Int.\ J.\ Geom.\ Meth.\ Mod.\ Phys.\  {\bf 13}, 1630002 (2016).



\bibitem{Busti:2015xqa} 
  V.~C.~Busti, Á.~de la Cruz-Dombriz, P.~K.~S.~Dunsby and D.~Sáez-Gómez,
  Phys.\ Rev.\ D {\bf 92}, 123512 (2015).








\bibitem{Dunajski:2008tg} 
  M.~Dunajski and G.~Gibbons,
  Class.\ Quant.\ Grav.\  {\bf 25}, 235012 (2008).




\bibitem{Rapetti:2006fv} 
  D.~Rapetti, S.~W.~Allen, M.~A.~Amin and R.~D.~Blandford,
  Mon.\ Not.\ Roy.\ Astron.\ Soc.\  {\bf 375}, 1510 (2007).

\bibitem{Zhai:2013fxa} 
  Z.~X.~Zhai, M.~J.~Zhang, Z.~S.~Zhang, X.~M.~Liu and T.~J.~Zhang,
  Phys.\ Lett.\ B {\bf 727}, 8 (2013).




\bibitem{Mukherjee:2016shl} 
  A.~Mukherjee and N.~Banerjee,
  Class.\ Quant.\ Grav.\  {\bf 34}, 035016 (2017).


\bibitem{Mukherjee:2016trt} 
  A.~Mukherjee and N.~Banerjee,
  Phys.\ Rev.\ D {\bf 93}, 043002 (2016).


\bibitem{Balcerzak:2014rga} 
  A.~Balcerzak and M.~P.~Dabrowski,
  JCAP {\bf 1406}, 035 (2014).







\bibitem{Aviles:2016wel} 
  A.~Aviles, J.~Klapp and O.~Luongo,
  Phys.\ Dark Univ.\  {\bf 17}, 25 (2017).
  




\bibitem{Alam:2003sc} 
  U.~Alam, V.~Sahni, T.~D.~Saini and A.~A.~Starobinsky,
  Mon.\ Not.\ Roy.\ Astron.\ Soc.\  {\bf 344}, 1057 (2003);

  V.~Sahni, T.~D.~Saini, A.~A.~Starobinsky and U.~Alam,
  JETP Lett.\  {\bf 77}, 201 (2003)
  [Pisma Zh.\ Eksp.\ Teor.\ Fiz.\  {\bf 77}, 249 (2003)].




\bibitem{Sahni:2008xx} 
  V.~Sahni, A.~Shafieloo and A.~A.~Starobinsky,
  Phys.\ Rev.\ D {\bf 78}, 103502 (2008).

\bibitem{Zunckel:2008ti} 
  C.~Zunckel and C.~Clarkson,
  Phys.\ Rev.\ Lett.\  {\bf 101}, 181301 (2008).



\bibitem{Capozziello:2018jya} 
  S.~Capozziello, Ruchika and A.~A.~Sen,
  arXiv:1806.03943 [astro-ph.CO].



\bibitem{Haridasu:2018gqm} 
  B.~S.~Haridasu, V.~V.~Luković, M.~Moresco and N.~Vittorio,
  JCAP {\bf 1810}, no. 10, 015 (2018).





\bibitem{Arabsalmani:2011fz} 
  M.~Arabsalmani and V.~Sahni,
  Phys.\ Rev.\ D {\bf 83}, 043501 (2011).







\bibitem{jla}M. Betoule {\it et al.}, Astron. Astrophys. {\bf 568}, A22 (2014).

\bibitem{ohdcc}C. Zhang, H. Zhang, S. Yuan, T.J. Zhang and Y.C. Sun, Res. Astron. Astrophys. {\bf 14}, 1221 (2014);\\
               J. Simon, L. Verde and R. Jimenez,Phys. Rev. D {\bf 71}, 123001 (2005);\\
               M. Moresco, L. Verde, L. Pozzetti, R. Jimenez and A. Cimatti, JCAP {\bf 07}(2012)053;\\
               M. Moresco, L, Pozzetti, A. Cimatti {\it et al.}  JCAP, 05(2016)014;\\ 
               A. L. Ratsimbazafy, S. I. Loubser, S. M. Crawford, {\it et al.} Mon.\ Not.\ Roy.\ Astron.\ Soc.\ {\bf 467}, 3239 (2017);\\ 
               M. Moresco, Mon.\ Not.\ Roy.\ Astron.\ Soc.\ {\bf 450}, L16 (2015).\\ 

\bibitem{ohdbao}S. Alam {\it et al.} Mon.\ Not.\ Roy.\ Astron.\ Soc.\ {\bf 470}, 2617 (2017);\\

\bibitem{ohdLya}T. Delubac {\it et al.},  Astron. Astrophys. {\bf 574} A59 (2015);\\
                A. Font-Ribera {\it et al.}, JCAP 05(2014)027.



\bibitem{6dF}F. Beutler {\it et al.}, Mon.\ Not.\ Roy.\ Astron.\ Soc.\ {\bf 416}, 3017 (2011).

\bibitem{Ross:2014qpa} 
  A.~J.~Ross, L.~Samushia, C.~Howlett, W.~J.~Percival, A.~Burden and M.~Manera,
  Mon.\ Not.\ Roy.\ Astron.\ Soc.\  {\bf 449}, 835 (2015).



\bibitem{Anderson:2013zyy} 
  L.~Anderson {\it et al.} [BOSS Collaboration],
  Mon.\ Not.\ Roy.\ Astron.\ Soc.\  {\bf 441}, 24 (2014).



\bibitem{emcee1} J. Goodman and J. Weare, Communication in Applied Mathematics and Computational Science {\bf 5}, 65 (2010).

\bibitem{emcee}D. Foreman-Mackey, D. W. Hogg, D. Lang and J. Goodman, Publ. Astron. Soc. Pac. {\bf 125}, 306 (2013).






\bibitem{Shafieloo2012}A. Shafieloo, V. Sahni, A. A. Starobinsky, Phys. Rev. D {\bf 86}, 103527 (2012).


\bibitem{numpy}S. V. D. Walt, S. C. Colbert, G. Varoquaux, Computing in Science and Engineering {\bf 13}, 22 (2011).
                 http://www.numpy.org/.

\bibitem{scipy}E. Jones, E. Oliphant, P. Peterson {\it et al.}, SciPy: Open Source Scientific Tools for Python,\\
               http://www.scipy.org/.


\bibitem{veusz} https://veusz.github.io/

\bibitem{Foreman-Mackey2016} D. Foreman-Mackey, “Corner.py on Github.” (2016).  \\
                   https://github.com/dfm/corner.py.



\bibitem{Yu:2017iju} 
  H.~Yu, B.~Ratra and F.~Y.~Wang,
  Astrophys.\ J.\  {\bf 856}, no. 1, 3 (2018).



\bibitem{Park:2018tgj} 
  C.~G.~Park and B.~Ratra,
  arXiv:1809.03598 [astro-ph.CO].








\bibitem{Ade:2015xua} 
  P.~A.~R.~Ade {\it et al.} [Planck Collaboration],
  Astron.\ Astrophys.\  {\bf 594}, A13 (2016).

\bibitem{Macaulay:2018fxi} 
  E.~Macaulay {\it et al.} [DES Collaboration],
  [arXiv:1811.02376 [astro-ph.CO]].



\end{thebibliography}
\end{document}